\documentclass[printer]{aa}

\usepackage{txfonts}
\usepackage{amsfonts}
\usepackage{tablefootnote}
\usepackage{graphicx}
\usepackage{lscape}
\usepackage{multirow}
\usepackage{dirtytalk}
\usepackage{framed,enumitem}
\setlist[itemize]{labelindent=0pt,labelwidth=0pt, labelsep*=5pt, leftmargin=0pt, style=standard}
\usepackage[nopostdot,acronym,xindy,indexonlyfirst=true,stylemods=longbooktabs,automake]{glossaries-extra}
\makeglossaries
\newacronym{ac}{AC}{across scan}
\newacronym{aen}{AEN}{astrometric excess noise}
\newacronym{al}{AL}{along scan}
\newacronym{bd}{BD}{brown dwarf}
\newacronym{cds}{CDS}{Centre de Données de Strasbourg}
\newacronym{cu8}{CU8}{Coordination unit 8}
\newacronym{dec}{DEC}{declination}
\newacronym{dof}{DOF}{degree of freedom}
\newacronym{dpac}{DPAC}{Data processing and analysis consortium}
\newacronym{dr}{DR}{data release}
\newacronym{flame}{FLAME}{final luminosity age and mass estimate}
\newacronym{fov}{FoV}{field of view}
\newacronym{fp}{FP}{false-positives}
\newacronym{gaston}{GASTON}{\textit{Gaia} astrometric noise simulation to derive orbital inclination}
\newacronym{gost}{GOST}{\textit{Gaia} observation forecast tool}
\newacronym{g3}{GDR3}{third \textit{Gaia} data release}
\newacronym{dr4}{GDR4}{fourth \textit{Gaia} data release}
\newacronym{hg}{HG}{\textsc{Hipparcos}--\textit{Gaia}}
\newacronym{hg3}{HG}{\textsc{Hipparcos} -- \textit{Gaia} DR3}
\newacronym{iads}{IADs}{intermediate astrometric data}
\newacronym{icrs}{ICRS}{International celestial reference system}
\newacronym{ipd}{IPD}{Image parameter determination}
\newacronym{lr}{LR}{likelihood ratio}
\newacronym{lsf}{LSF}{line spread function}
\newacronym{mad}{MAD}{median absolute deviation}
\newacronym{mcmc}{MCMC}{Markov-chains Monte-Carlo}
\newacronym{ms}{MS}{main-sequence}
\newacronym{mse}{UEVA}{unbiased estimator of variance a posteriori}
\newacronym{nea}{NEA}{NASA Exoplanet Archive}
\newacronym{nss}{\textit{Gaia}-NSS}{\textit{Gaia} DR3 non single star catalog}
\newacronym{pa}{PA}{position angle}
\newacronym{pdf}{PDF}{probability density function}
\newacronym{pma}{PMa}{proper motion anomaly}
\newacronym{pmex}{GaiaPMEX}{\textit{Gaia} DR3 proper motion anomaly and astrometric noise excess}
\newacronym{psf}{PSF}{point spread function}
\newacronym{ra}{RA}{right ascension}
\newacronym{rse}{RSE}{regression standard error}
\newacronym{rms}{RMS}{root mean square}
\newacronym{rss}{RSS}{residuals sum of square}
\newacronym{ruwe}{\texttt{ruwe}}{renormalised unit weight error}
\newacronym{rv}{RV}{radial velocity}
\newacronym{sma}{sma}{semi-major axis relative to the central star}
\newacronym{snr}{S/N}{signal-to-noise ratio}
\newacronym{uwe}{UWE}{unit weight error}
\newacronym{wc}{WC}{window class}
\newacronym{wfs}{WFS}{wavefront sensor}
\newacronym{5p}{5p}{5-parameters}
\newacronym{6p}{6p}{6-parameters}

\makeatletter
\renewcommand*\aa@pageof{, page \thepage{} of \pageref*{LastPage}}
\makeatother

\usepackage[breaklinks,pdfencoding=auto,psdextra]{hyperref}
\usepackage{color}

\begin{document}

\title{Searching for substellar companion candidates with \textit{Gaia}}
\subtitle{II. A catalog of 9,698 planet candidate solar-type hosts\thanks{Table B.1 is only available in electronic form at the CDS via anonymous ftp to cdsarc.u-strasbg.fr (130.79.128.5) or via http://cdsweb.u-strasbg.fr/cgi-bin/qcat?J/A+A/}}
\author{F. Kiefer\inst{1} \and A.-M. Lagrange\inst{1}  \and P. Rubini\inst{2} \and F. Philipot\inst{1}}

\institute{
\label{inst:1}LESIA, Observatoire de Paris, Universit\'e PSL, CNRS, Sorbonne Universit\'e, Universit\'e de Paris, 5 place Jules Janssen, 92195 Meudon, France\thanks{Please send any request to flavien.kiefer@obspm.fr} \and
\label{inst:2}Pixyl, 5 av du Grand Sablon 38700 La Tronche
}
        
   \date{Received 31/07/2024 ; accepted 03/09/2024}

  
  \abstract 
{In a previous paper, we introduced a new tool called "\glsxtrlong{pmex}", or \glsxtrshort{pmex}. This tool characterizes the mass and \glsxtrlong{sma} (\glsxtrshort{sma}) of a possible companion around any source observed with \textit{Gaia} using the value of renormalized unit weight error (\glsxtrshort{ruwe}), or with both \textit{Gaia} and \textsc{Hipparcos} using the value of the \glsxtrlong{pma} (\glsxtrshort{pma}), either alone or combined with the \glsxtrshort{ruwe}. 
}
{Our goal is to exploit the large volume of sources in the \glsxtrlong{g3} catalog to find new exoplanet candidates. We wish to create a new input catalog of planet-candidate-hosting systems to enable future follow-up projects. Beyond magnitude 14, this catalog would prepare the arrival of powerful instruments on the Extremely Large Telescopes, which could include radial velocity (RV) follow-up of faint stars and direct imaging of planets around main sequence stars of gigayear ages. }
{We used the mass--\glsxtrshort{sma} degenerate set of solutions obtained by \glsxtrshort{pmex} from any value of \glsxtrshort{ruwe} to select a sample of bright ($G$$<$16) \textit{Gaia} sources whose companions could be in the planetary domain, with a mass of $<$13.5\,M$_{\rm J}$. We selected sources whose astrometric signature determined from the \glsxtrshort{ruwe} is larger than zero with a significance of $>$2.7--$\sigma$ ($p$-value $<$0.00694).}
{It led us to identify a sample of 9,698 planet-candidate-hosting sources that have a companion with a mass of possibly $<$13.5\,M$_{\rm J}$ and a \glsxtrshort{sma} in the
range of $\sim$1--3\,au. 
We cross-matched our catalog with the \glsxtrlong{nea} (\glsxtrshort{nea}) catalog of exoplanets, identifing 19 of our systems therein. We successfully detected eight confirmed substellar companions with an sma of 1--3\,au, initially discovered and characterized with RV and astrometry. Moreover, we found six transiting-planet systems and two wide-orbit systems for whom, with \glsxtrshort{pmex}, we predict the existence of supplementary companions. Focusing on the subsample of sources observed with \textsc{Hipparcos}, combining the constraints from \glsxtrshort{ruwe} and \glsxtrshort{pma}, we confirm the identification of four new planetary candidate systems, HD 187129, HD 81697, CD-42 883, and HD 105330.}
{Given the degeneracy of mass--\glsxtrshort{sma}, many of the candidates in this catalog of 9,698 sources might have a larger mass in the brown-dwarf and stellar domain if their \glsxtrshort{sma} departs from the 1--3-au range. The vetting of this large catalog will be the subject of future studies.}


   \maketitle
%

\section{Introduction}

Up to now, the vast majority of exoplanets have been discovered using the transit and \glsxtrlong{rv} (\glsxtrshort{rv}\footnote{All acronyms used are summarized and indexed in Appendix~\ref{sec:acronyms}.}) techniques, as seen for example in the \glsxtrlong{nea}\footnote{\url{https://science.nasa.gov/exoplanets/exoplanet-catalog/}} (\glsxtrshort{nea}) or the \texttt{exoplanet.eu} catalog. \textit{Gaia}’s absolute astrometry is expected to reveal (tens of) thousands of new exoplanets and brown dwarfs (BDs) in the near future~\citep{Perryman2014,Sahlmann2015,Holl2022,Arenou2023,Holl2023}. 
Using the \glsxtrlong{g3} (\glsxtrshort{g3};~\citealt{Gaia2021}) time series,~\citet{Holl2023} recently published 1162 sources with an astrometric-orbit solution, including 9 exoplanet candidates and 29 \glsxtrshort{bd} candidates (assuming an $M_\star$ of 1\,M$_\odot$). Those candidates have masses whose confidence region overlaps with the planetary or \glsxtrshort{bd} domain. Based on the Thiele-Innes parameters fitted from the unpublished time series and listed in the \glsxtrlong{nss} (\glsxtrshort{nss}),~\citet{Arenou2023} reported 1843 \glsxtrshort{bd} candidates and 72 exoplanet candidates, among which there are 10 already known\,\glsxtrshort{bd}s, and 9 \glsxtrshort{rv} exoplanets validated with \textit{Gaia}'s astrometry, and 2 new exoplanets with masses of 5 and 7\,M$_{\rm J}$ , which were also identified in~\citet{Holl2023}.

Given the many thousands of exoplanets expected from \textit{Gaia}, the above-reported number of exoplanet candidates is still below expectations. This could be partly explained by the sparse temporal coverage of orbital phases, which leads to large uncertainties and degeneracies on the orbital solutions. 
Nevertheless, \textit{Gaia} should allow the detection of numerous Jupiter-mass exoplanets within the range of Earth--Neptune orbits. This range is still underpopulated among the approximately 5\,000 known exoplanets because of the observation biases inherent to the techniques that yield many of the detections: short-period ($P$$<$1\,yr) planets are mainly detected and characterised with \glsxtrshort{rv} and transits, while high-contrast imaging is most sensitive to super-massive exoplanets and those with a large semi-major axis ($M_p$$>$5\,M$_{\rm J}$; \glsxtrshort{sma}$>$5\,au). A key objective is the exploitation of the \textit{Gaia} database in its most recent release (\glsxtrshort{g3}; ~\citealt{Gaia2021}) in order to detect unknown exoplanet candidates, such as AF\,Lep\,b~\citep{Mesa2023,Franson2023,DeRosa2023}.  

In Kiefer et al. (2024; Paper I hereafter), we introduced a new tool called \glsxtrlong{pmex} (\glsxtrshort{pmex}), which is designed to determine the mass of possible candidate  companions and their  sma relative to their central star   by considering either the \glsxtrlong{aen} (\glsxtrshort{aen}; see \citealt{Kiefer2019a,Kiefer2019b,Kiefer2021}), the renormalized unit weight error (\glsxtrshort{ruwe}; see~\citealt{Lindegren2018,Lindegren2021}), or the \glsxtrlong{pma} (\glsxtrshort{pma}; see \citealt{Kervella2019,Brandt2021,Kervella2022}), or by combining the constraints from the \glsxtrshort{ruwe} and the \glsxtrshort{pma}. 

In Sect.~\ref{sec:pmex} we briefly describe the theory behind \glsxtrshort{pmex} and the main properties of the mass--sma solutions found that led us to determine a minimum companion mass compatible with \textit{Gaia}'s astrometry 
for any source brighter than $G$=16. In Sect.~\ref{sec:catalog} we present a sample of 9,698 sources around which we infer the presence of a companion whose mass could overlap with the planetary domain. This catalog is discussed and compared with other catalogs in Sect.~\ref{sec:discussion}.

\section{The astrometric minimum mass of companions determined with GaiaPMEX}
\label{sec:pmex}
The \glsxtrshort{pmex} tool models ---within a Bayesian framework--- the \glsxtrshort{aen}, \glsxtrshort{ruwe}, and \glsxtrshort{pma} as measured by \textit{Gaia} and \textsc{Hipparcos} based on the orbital motion of a source's photocenter due to a companion, and accounts for both measurement and instrumental noise. GaiaPMEX provides 2D confidence maps for the mass and sma of a companion based on the individual values of \glsxtrshort{aen}, \glsxtrshort{ruwe}, and \glsxtrshort{pma}, and on the constraints from \glsxtrshort{ruwe} and \glsxtrshort{pma} combined. 

An interesting feature of these confidence maps is that they follow a specific pattern; the maps determined from the \glsxtrshort{aen} and \glsxtrshort{ruwe} are V-shaped, while those determined from the \glsxtrshort{pma} are U-shaped. This pattern is well modeled by segmented linear relationships between mass and \glsxtrshort{sma}, which are thoroughly detailed in Paper I. These relationships depend on the part of the \glsxtrshort{aen}, \glsxtrshort{ruwe}, or \glsxtrshort{pma} in excess of noise, which we call the "astrometric signature". 
\begin{figure}[hbt]
    \centering
    \includegraphics[width=89.3mm,clip=true]{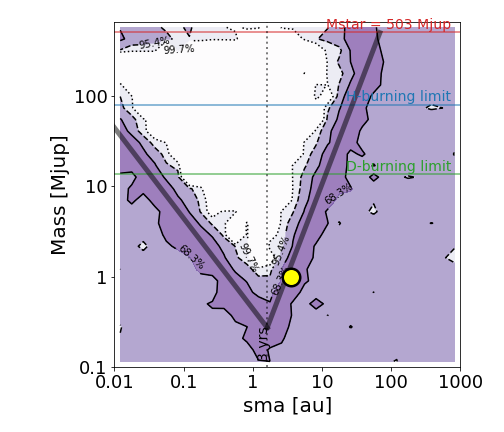}
    \caption{\glsxtrshort{pmex} confidence map of the mass and sma of a companion around GJ\,832 constrained by the \glsxtrshort{ruwe}. The darkest area delineated with a black solid line spans the 68.3\% confidence region. The gradually lighter purple areas delineated with black dashed and dotted lines,   respectively, span the 95.4\% and 99.7\% confidence regions. This example is described in detail in Paper I. The thick dark lines show the mass--\glsxtrshort{sma} relationship of Eq.~\ref{eq:mass_sma_AEN}. The yellow circle shows the $M\sin i$ and sma of the known Jupiter-like planet in this system~\citep{Philipot2023b}.}
    \label{fig:GJ832_map}
\end{figure}
We are particularly interested in the relationships fixed by the value of the \glsxtrshort{ruwe}; these are shown in Fig.~\ref{fig:GJ832_map} for the illustrative case of the M-dwarf GJ\,832 (see also Paper I for more details). 

\subsection{The \texttt{ruwe}-based astrometric signature}
\label{sec:alpha_mse}
The \glsxtrshort{ruwe} measures the amplitude of residuals relative to the formal errors of the data, once a five-parameter linear model (including centroid position, proper motion, and parallax) has been fitted out of \textit{Gaia}'s astrometric points. From the \glsxtrshort{ruwe}, we are able to determine a \glsxtrshort{ruwe}-based astrometric signature, referred to here as $\alpha_{\rm \glsxtrshort{mse},\glsxtrshort{ruwe}}$. The astrometric signature $\alpha_{\rm \glsxtrshort{mse}}$ measures the excess in the \glsxtrlong{mse} (\glsxtrshort{mse} for short) of the five-parameter model residuals, beyond the expectation value of the \glsxtrlong{mse} if the star was actually single: 
\begin{equation}
    \alpha_{\rm \glsxtrshort{mse}} = \sqrt{{\rm \glsxtrshort{mse}} - {\rm \glsxtrshort{mse}}_{\rm single}}
.\end{equation}

The \glsxtrshort{mse} of a given source can be determined from the value of \glsxtrshort{ruwe}\footnote{It could also be determined from the \glsxtrshort{aen}, but as explained in paper\,I, the \glsxtrshort{ruwe} is more reliable for all sources of any $G$$<$16.} using the equation derived in Paper I, that is, 
\begin{equation}
{\rm \glsxtrshort{mse}}_{\rm ruwe}=\left({\rm \glsxtrshort{ruwe}} \times u_0\right)^2 \, \left(\sigma^2_{\rm att}+\sigma^2_{\rm AL}\right) \label{eq:ruwe_mse}
,\end{equation}
where $u_0$ is the factor determined for any source with respect to its $G$-mag and $Bp-Rp$ in the \glsxtrshort{g3} auxiliary data\footnote{\url{https://www.cosmos.esa.int/web/gaia/auxiliary-data}}. 
The $\sigma_{\rm \glsxtrshort{al}}$ is the error on the astrometric \glsxtrlong{al} (\glsxtrshort{al}) angle measurements, and $\sigma_{\rm att}$ is the attitude excess noise, that measures an 'average' calibration noise $\sim$0.076\,mas among all sources observed at a given epoch. 

The single star \glsxtrshort{mse}$_{\rm single}$ is determined from the theoretical approximate normal distribution of the \glsxtrshort{mse} with respect to typical errors and noise in \glsxtrshort{g3} data found in Paper I, ${\mathcal N}\left(\mu,\sigma\right)$: 
\begin{align}
    \mu =& \frac{N_{\glsxtrshort{al}}}{N_{\glsxtrshort{al}}\,N_{\rm \glsxtrshort{fov}} - 5}\,\left[(N_{\rm \glsxtrshort{fov}}-5) \, \sigma^2_{\rm calib} + N_{\rm \glsxtrshort{fov}}\,\sigma^2_{\rm AL}\right] \label{eq:mse_single} \\ 
    \sigma^2=&  \frac{2 N_{\glsxtrshort{al}}}{\left(N_{\glsxtrshort{al}}\,N_{\rm \glsxtrshort{fov}} - 5\right)^2}\,\Bigg[ N_{\glsxtrshort{al}} \left(N_{\rm \glsxtrshort{fov}}-5\right)\,\sigma^4_{\rm calib} \nonumber \\ & \qquad\qquad\qquad +  N_{\rm \glsxtrshort{fov}}\,\sigma^4_{\rm AL}  + 2\,N_{\rm \glsxtrshort{fov}} \,\sigma_{\rm AL}^2\,\sigma_{\rm calib}^2\Bigg], \label{eq:mse_single_sigma}
\end{align}
where $\sigma_{\rm calib}$ is the calibration noise. The values of $\sigma_{\rm calib}$, $\sigma_{\rm \glsxtrshort{al}}$ and $\sigma_{\rm att}$ are estimated in Paper I, with respect to the $G$, $Bp-Rp$, \glsxtrshort{ra}, and Dec. of any given sources. $N_{\rm \glsxtrshort{fov}}$ is the number of \glsxtrlong{fov} (\glsxtrshort{fov}) transits detected by \textit{Gaia} (\verb+astrometric_matched_transit+) and $N_{\glsxtrshort{al}}$ is the average number of astrometric measurement per transit $N_{\glsxtrshort{al}}$=${\rm int}(N/N_{\rm \glsxtrshort{fov}})$, with $N$ the total number of \glsxtrshort{al} angle measurements (\verb+astrometric_n_good_obs_al+).

\subsection{The significance of \texorpdfstring{$\alpha_{\rm UEVA,\texttt{ruwe}}$}{the \texttt{ruwe}-based astrometric signature}}
\label{sec:significance}
Even if $\alpha_{\rm \glsxtrshort{mse},\glsxtrshort{ruwe}}$ is above zero, it remains possible that it only stems from the effect of astrometric noise when observing a single source. In such a case indeed, $\alpha_{\rm \glsxtrshort{mse},\glsxtrshort{ruwe}}$ would be said to not be significant. To express the probability for $\alpha_{\rm \glsxtrshort{mse},\glsxtrshort{ruwe}}$ to be explained by only noise, we define its significance as the $N-\sigma$ level that corresponds to the one-sided $p$-value of $\rm \glsxtrshort{mse}_{\rm \glsxtrshort{ruwe}}^{1/3}$ in the distribution of ${\rm \glsxtrshort{mse}}_{\rm single}^{1/3}$. The higher the significance, the lower the probability of a noise-based origin, and thus of singleness. The 1, 2, and 3--$\sigma$ levels correspond to $p$-values of respectively 31.7\%, 4.6\%, and 0.3\%. As detailed in paper I, the reason for using the $\rm \glsxtrshort{mse}$ instead of $\alpha_{\rm \glsxtrshort{mse},\glsxtrshort{ruwe}}$ directly, and to the power $1/3$, is that (i) $\alpha_{\rm \glsxtrshort{mse}}$ is undefined whenever ${\rm \glsxtrshort{mse}}_{\rm \glsxtrshort{ruwe}}$ is smaller than ${\rm \glsxtrshort{mse}}_{\rm single}$, and (ii) according to~\citet{Wilson1931} (see also~\citealt{canal2005}), if $X$ is proportional to a random variable that follows a $\chi^2$ distribution, then $X^{1/3}$ closely follows a normal distribution. This is approximately the case for ${\rm \glsxtrshort{mse}}_{\rm single}$ (see Sects. 5.2.1 and C in paper I). We therefore assumed that \glsxtrshort{mse}$^{1/3}$ followed a normal distribution ${\mathcal N} \left(\mu_{1/3},\sigma_{1/3}\right)$ and approximated its parameters by $\mu_{1/3}$=$\mu^{1/3}$ and $\sigma_{1/3}$=$ \sigma\,\mu^{-2/3} /3$ by applying error propagation from Eqs.~\ref{eq:mse_single} and~\ref{eq:mse_single_sigma}. It is then straightforward to calculate a $p$-value for a realisation $x$ using the \verb+Python+ function \texttt{scipy.stats.normal.cdf(x,$\mu_{1/3}$,$\sigma_{1/3}$)} and a corresponding $N$--$\sigma$ using \texttt{scipy.stats.normal.ppf($1-(1-p)/2$)}.

\subsection{The mass--sma relationships and the minimum mass}
The astrometric signature $\alpha_{\rm \glsxtrshort{mse}}$ relates the \glsxtrshort{sma} and the mass through a formula determined in Paper I, in which the parameters vary with the orbital period of the companion:

\begin{equation}
    M_c = C_{\ell} \, \frac{\alpha_{\rm \glsxtrshort{mse}} }{\varpi} \,M_\star^{\frac{2-n_\ell}{3}} \, {\rm sma}^{n_\ell} \label{eq:relation_summarized}
,\end{equation}
where the parameters $C_\ell$ and $n_\ell$ are fixed to
\begin{align}
\alpha_{\rm \glsxtrshort{mse}} & \Longrightarrow
    \begin{cases}
        \text{$\ell$=1: $<$3\,yr,} & n_1=-1\text{,}~~C_1=2300 \\
        \text{$\ell$=2: $>$3\,yr,} & n_2=+2\text{,}~~C_2=260.
    \end{cases} \label{eq:mass_sma_AEN}  \\ \nonumber
\end{align}
Those relationships are overplotted on the \glsxtrshort{pmex} confidence map determined from the \glsxtrshort{ruwe} of GJ\,832 in Fig.~\ref{fig:GJ832_map}. If the $\alpha_{\rm \glsxtrshort{mse},\glsxtrshort{ruwe}}$ is  significant, which means typically more than 2--$\sigma$, these relationships allow us to measure the minimum mass of the companion, located at the minimum of the V-shaped curve. This is reached at an \glsxtrshort{sma}$\sim$2\,au: 
\begin{align}
\alpha_{\rm \glsxtrshort{mse},\glsxtrshort{ruwe}} & \Longrightarrow
    \begin{cases}
        M_{c,\text{min}} = 1150 \,M_\star^{2/3} \times \alpha_{\rm \glsxtrshort{mse},\glsxtrshort{ruwe}}/\varpi \quad (\text{M}_{\rm J}),\\
        \text{\glsxtrshort{sma}}_{\rm min} = 2.1\,M_\star^{1/3} \quad (\text{au}).
    \end{cases} \label{eq:mass_minimum_AEN} \\ \nonumber 
\end{align}
We use this approximative $M_{c,\text{min}}$ to select a sample of candidate sources that may host a companion down to the planetary domain, whose $M_{c,\text{min}}$$<$13.5\,M$_{\rm J}$.

\section{A catalog of 9,698 stars with possible exoplanet candidates identified with \textit{Gaia}}
\label{sec:catalog}

\subsection{The main steps of the catalog selection}
\begin{figure}[hbt]
    \centering
    \includegraphics[width=85.3mm]{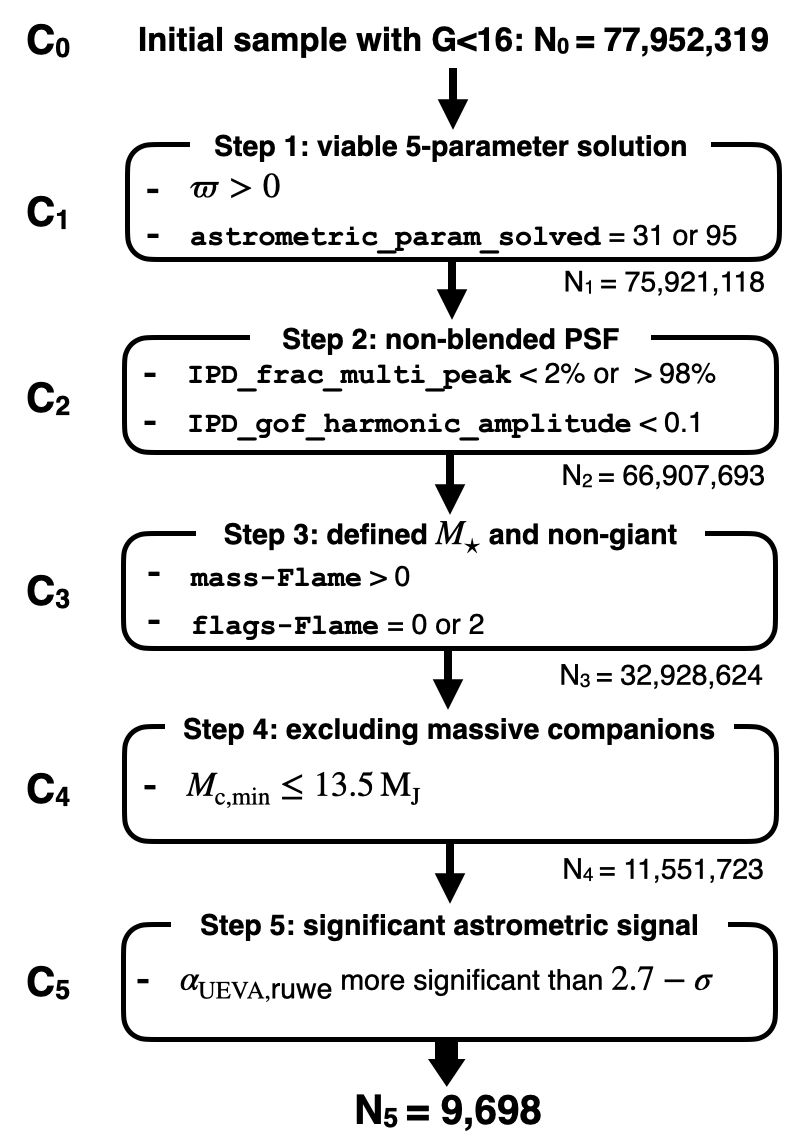}
    \caption{Selection steps used to build the sample of candidate systems with exoplanet companions, where six selection criteria ($C_0$ to $C_5$) are applied  iteratively.}
    \label{fig:steps}
\end{figure}

We searched the \glsxtrshort{g3} catalog for bright sources with $G$$<$16 that show a significant astrometric signature $\alpha_{\rm UEVA}$ compatible with a companion mass of $<$13.5\,M$_{\rm J}$. The diagram in Fig.~\ref{fig:steps} summarizes the various steps of our selection process. 

As we developed \glsxtrshort{pmex} specifically for bright sources with $G$$<$16 (Paper I), we used here the same parent sample of 77,952,319 sources obtained using this selection criterion, that is $G<16$. This is our input sample selection criterion $C_0$.

In step 1, we selected all sources with a positive parallax\footnote{Few sources have an ill-defined negative or null parallax. Those are discarded.} in the \glsxtrshort{5p} and \glsxtrshort{6p} datasets (criterion $C_1$), which gives a total of 75,921,118 sources, of which  73,582,079 belong to the 5p dataset and 2,339,039 to the 6p dataset. In step 2, applying criterion $C_2$, we rejected the sources that show diagnostics of blend in the \glsxtrshort{psf} fitted by the image parameter determination (\glsxtrshort{ipd}). As defined in Paper I (see also~\citealt{Fabricius2021}), strong blends in the \glsxtrshort{psf} due to background objects or other multiple components are diagnosed by two \glsxtrshort{ipd} indicators published in the \glsxtrshort{g3} archive. When \verb+IPD_frac_multi_peak+ is typically within 2--98\% and \verb+IPD_gof_harmonic_amplitude+$>$0.1, the centroid of the fitted \glsxtrshort{psf} likely underwent strong time-dependent offsets leading to significant residual signals on top of the proper and parallactic motions. Removing the sources satisfying these criteria led to the selection of 66,907,693 sources (65,677,199 in the 5p dataset and 1,230,494 in the 6p dataset). 

\begin{figure}
    \centering
    \includegraphics[width=89.3mm,clip=true]{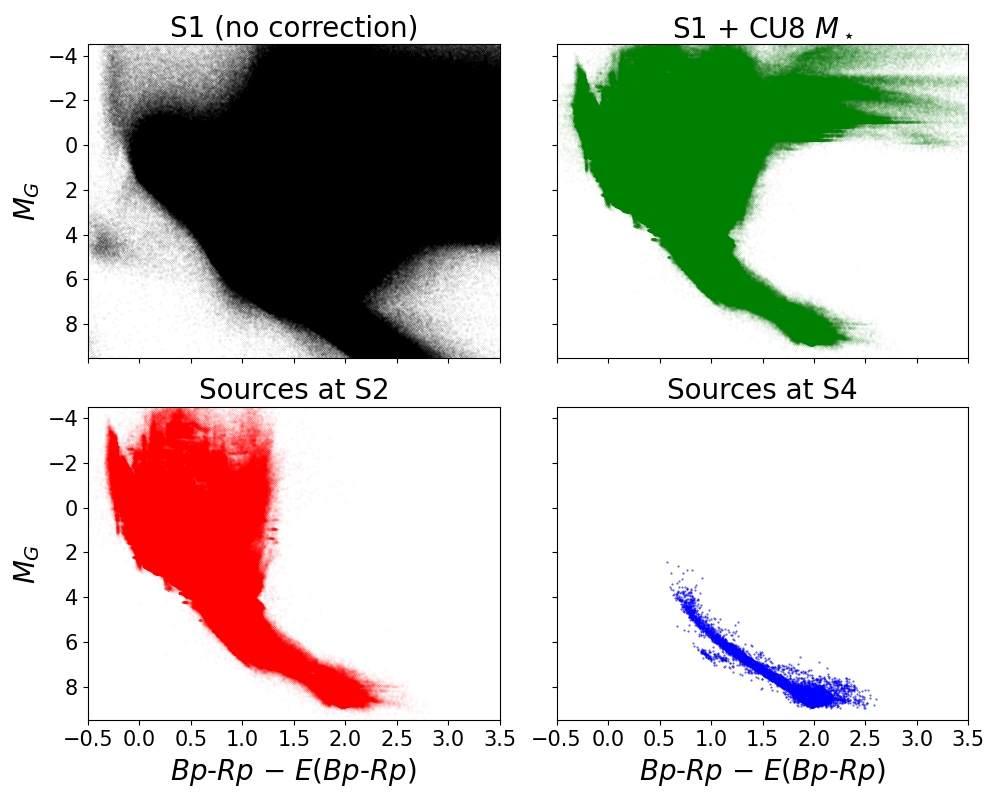}
    \caption{Hertzsprung-Russell diagrams throughout the sequence of the selection steps presented in Fig.~\ref{fig:steps}. The different steps are shown as follows: the sources at S1 whose absolute magnitude $M_G$ is not corrected for extinction and $Bp-Rp$ is not corrected for reddening are shown in black; the sources with an existing mass in the CU8 catalog and whose $G$ and $Bp-Rp$ are corrected for extinction and reddening are shown in green, and those that are moreover nongiant and without photometric mass are shown in red; and the final 9,698 planet-candidate hosts are shown in blue. }
    \label{fig:HRdiagram}
\end{figure}

In step 3, we applied our third selection criterion ($C_3$). We identified the sources with a well-defined and nongiant stellar mass in the \glsxtrshort{g3} \glsxtrlong{cu8} database (\glsxtrshort{cu8}; \citealt{Creevey2023}). Only about 140 million of the 1.6 billion stars in the full \glsxtrshort{g3} catalog have a proposed stellar mass of within 0.5--10\,M$_\odot$ as determined by combining photometry, parallax, and stellar models, the so-called \texttt{mass-Flame} in the CU8 database, which is also written $M_{\star,{\rm Flame}}$. A flag called \texttt{flags-Flame}, a two-characters string AB where A and B are either "0", "1", or "2", indicates the quality of the mass estimation in \glsxtrshort{cu8} (see details in \citealt{Creevey2023}. We excluded stars identified as giants (A=1 in \texttt+flags-Flame+), and those for which the distance was estimated photometrically (by GSP-phot; B=1 in \texttt+flags-Flame+) and is thus possibly inaccurate given the binarity of the stars that we are considering. This led us to select 32,928,624 sources, among which 32,422,316 and 506,308 are in the \glsxtrshort{5p} and \glsxtrshort{6p} datasets, respectively. Figure~\ref{fig:HRdiagram} shows the Hertzsprung-Russell diagram of all these sources and how they are reduced when: selecting sources given a defined \glsxtrshort{cu8} $M_{\star,{\rm Flame}}$, and then excluding giants and masses determined using photometric distances. In this diagram, we use the absolute magnitude and $Bp-Rp$ color corrected for extinction and reddening for the sources whose mass was found in the \glsxtrshort{cu8} catalog; otherwise the absolute magnitude was obtained as $M_G=G-5\log \varpi + 10$. Those sources not found in the \glsxtrshort{cu8} catalog are of diverse types, including evolved states ---giants and white dwarfs--- whose masses are more difficult to estimate.

In step 4, using our knowledge of $M_{\star,{\rm Flame}}$, \glsxtrshort{ruwe}, and parallax ($\varpi$), as well as the noises derived from the $G$-mag, $Bp-Rp$, \glsxtrshort{ra}, and Dec., we determined the mass by minimizing the \glsxtrshort{pmex} curves for \glsxtrshort{ruwe} and $M_{\rm c,min}$ using Eq.~\ref{eq:mass_minimum_AEN}. As we are interested in the planet candidates, we selected only those with $M_{\rm c,min}$$<$13.5\,M$_{\rm J}$ (criterion $C_4$). This led us to select a total of 11,551,723 systems, among which 11,342,194 are in the \glsxtrshort{5p} dataset, and 209,529 in the \glsxtrshort{6p} dataset. 

\begin{figure}
    \centering
    \includegraphics[width=89.3mm]{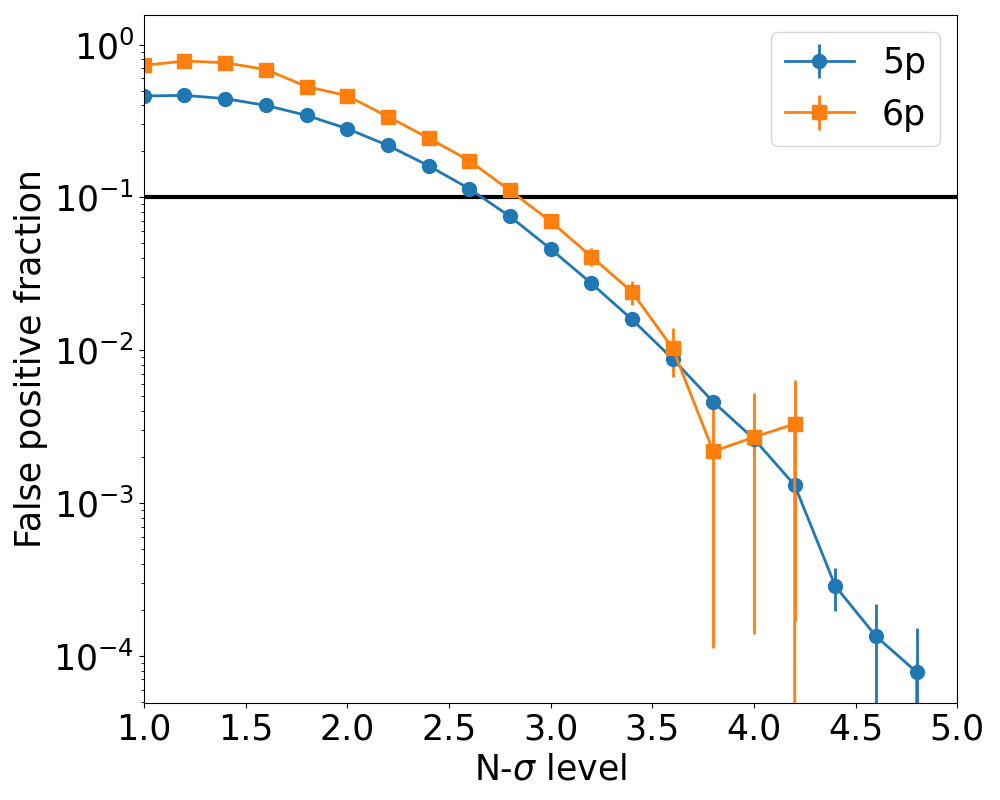}
    \caption{Fraction of FPs among selected planet candidates at different significance criteria ($N$--$\sigma$) in the 5p (blue) and 6p (orange) datasets. The black solid line shows the 10\% level.}
    \label{fig:FPratio}
\end{figure}

Finally, in step 5, we selected the sources for which, additionally, we have strong evidence of astrometric motion (criterion $C_5$). We selected the sources whose significance of $\alpha_{\rm \glsxtrshort{mse}}$ is greater than some threshold, $X$--$\sigma$, which optimizes the number of single-star  false positives (\glsxtrshort{fp}s) among the selected planet-candidate hosts sample. Those \glsxtrshort{fp}s would be single stars whose noise mimics the astrometric signature of stars with companions. The definition of $\alpha_{\rm \glsxtrshort{mse}}$ and its significance are described in Sects.~\ref{sec:alpha_mse} and \ref{sec:significance}. As we are interested in gathering as many candidates as possible, whilst keeping the number of \glsxtrshort{fp} single stars to 10\% of the selected sample
at most, we optimized the adopted significance threshold. The \glsxtrshort{fp} fraction at any $X$--$\sigma$ threshold is the number of \glsxtrshort{fp}s divided by the number of $\alpha_{\rm \glsxtrshort{mse}}$ that are more significant than $X$--$\sigma$ and lead to a companion mass of $<$13.5\,M$_{\rm J}$. We estimated this fraction using the following methodology. We took the sample defined in step 3 with $N_{\rm S3}$=32,928,624 sources, and simulated ---for all of these sources--- values of \glsxtrshort{mse} and $\alpha_{\rm \glsxtrshort{mse}}$ as if they were all single stars, that is, only considering stochastic astrometric variability. To do so, we drew values of \glsxtrshort{mse}$^{1/3}$ from the normal distribution ${\mathcal N} \left(\mu_{1/3},\sigma_{1/3}\right)$ defined in Sect.~\ref{sec:alpha_mse} (Eqs.~\ref{eq:mse_single} and~\ref{eq:mse_single_sigma}), from which we derived \glsxtrshort{mse} and $\alpha_{\rm \glsxtrshort{mse}}$. Using Eq.~\ref{eq:mass_minimum_AEN}, we calculated the corresponding minimum masses, and selected those of $<$13.5\,M$_{\rm J}$ (criterion $C_4$). Then, we selected the sources with \glsxtrshort{mse}$^{1/3}$ more significant than $X$--$\sigma$ (criterion $C_5$). This led to a \glsxtrshort{fp} planet sample at $X$ with a size of $N_{{\rm \glsxtrshort{fp}},0}$. However, since a large fraction of the 32,928,624 sources  are truly multiple, and thus not single, $N_{{\rm \glsxtrshort{fp}},0}$ overestimates the actual number of \glsxtrshort{fp}. A more realistic size of the single-star population in the input sample is obtained by (i) removing the sources with \glsxtrshort{mse}$_{\rm \glsxtrshort{ruwe}}^{1/3}$ larger than $X$--$\sigma$, and (ii) dividing the size, $N_{X-\sigma}$, of this residual sample by the theoretical proportion, $r_{X-\sigma}$, below $X$--$\sigma$ significance. For instance, $r_{X-\sigma}$ takes values of 0.683, 0.954, or 0.9973 if $X$=1, 2, or 3, respectively. This leads to $N_{\rm \glsxtrshort{fp}}=N_{{\rm \glsxtrshort{fp}},0}\times \left(N_{X-\sigma}/r_{X-\sigma}\right)/N_{\rm S3}$, of which the \glsxtrshort{fp} fraction among candidate-planet systems beyond $X$--$\sigma$ significance can be determined.
We tested many values of $X$--$\sigma$ significance from 1 to 5, and reproduced the whole process ten times to determine the mean and standard deviation of the FP fraction as a function of the significance level. This is shown in Fig.~\ref{fig:FPratio}. 
The FP fraction is smaller than 10\% when selecting planet candidates whose $\alpha_{\rm \glsxtrshort{mse},\glsxtrshort{ruwe}}$ is at least above the 2.7--$\sigma$ significance level, that is, with $p$-value$<$0.006934. Adopting this 2.7--$\sigma$ level led us to select a sample of 9,698 sources, among which 9,587 sources are in the 5p dataset, and 111 are in the 6p dataset. Being more selective on criterion $C_5$ leads to subsamples of 6,552 systems adopting a 3--$\sigma$ significance level ($p$-value$<$0.0027), and 2,268 with 4--$\sigma$ ($p$-value$<$0.000063). Those datasets would have a FP fraction of about 6\% and 0.2\%,  respectively. The full catalog of the 9,698 candidate sources is available at the CDS, with a small extract shown in Table~\ref{tab:planet_catalog}. We ordered the catalog in increasing magnitude in $G$-band. The source with ID \# 1 has the smallest magnitude, of 5.99, and the source with ID \# 9698 has the highest magnitude, of 15.53.

\subsection{Sample description}

The Hertzsprung-Russell diagram of the planet-candidate hosts sample is shown in Fig.~\ref{fig:HRdiagram}. While at step 3 some subgiants remain in the sample, all selected planet-candidate hosts at step 5 (in blue) are located along the main sequence. This is explained by the selection of sources with $M_{\star,{\rm Flame}}$ of mostly $<$1\,M$_\odot$. 

\begin{figure}[hbt]
    \centering
    \includegraphics[width=89.3mm]{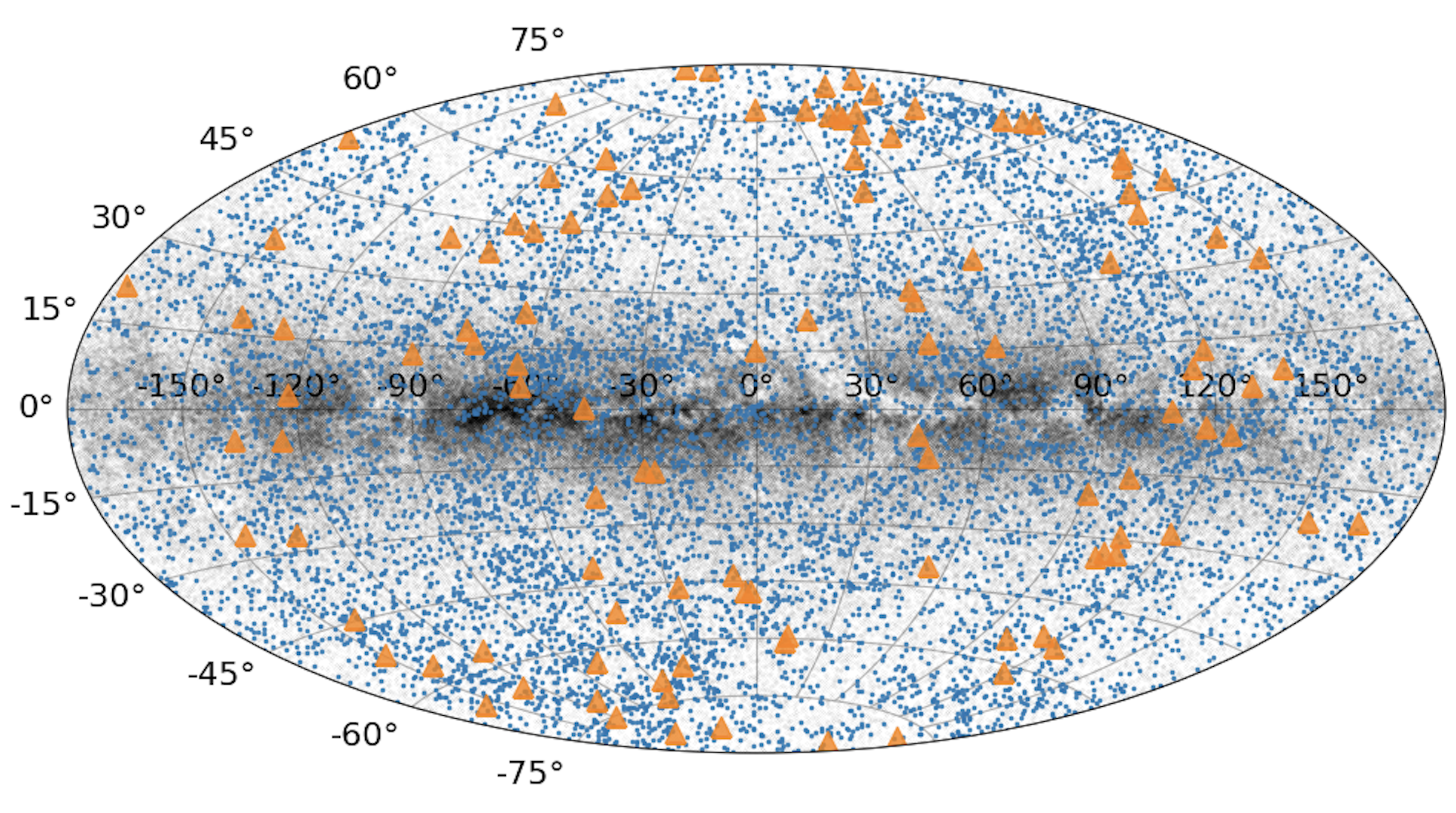}
    \caption{Distribution of the locations in the plane of the sky, in galactic coordinates, of planet-candidate hosts from the \glsxtrshort{5p} (blue) and \glsxtrshort{6p} (orange) datasets, compared to all the sources in the input sample at step 3 (black).}
    \label{fig:galactic_sky}
\end{figure}

Figure~\ref{fig:galactic_sky} shows the distribution of planet-candidate hosts across the sky. They are homogeneously distributed compared to the full input sample at step 3. The hosts are thus field stars that are relatively close to the Sun. 

\begin{figure}[hbt]
    \centering
    \includegraphics[width=89.3mm]{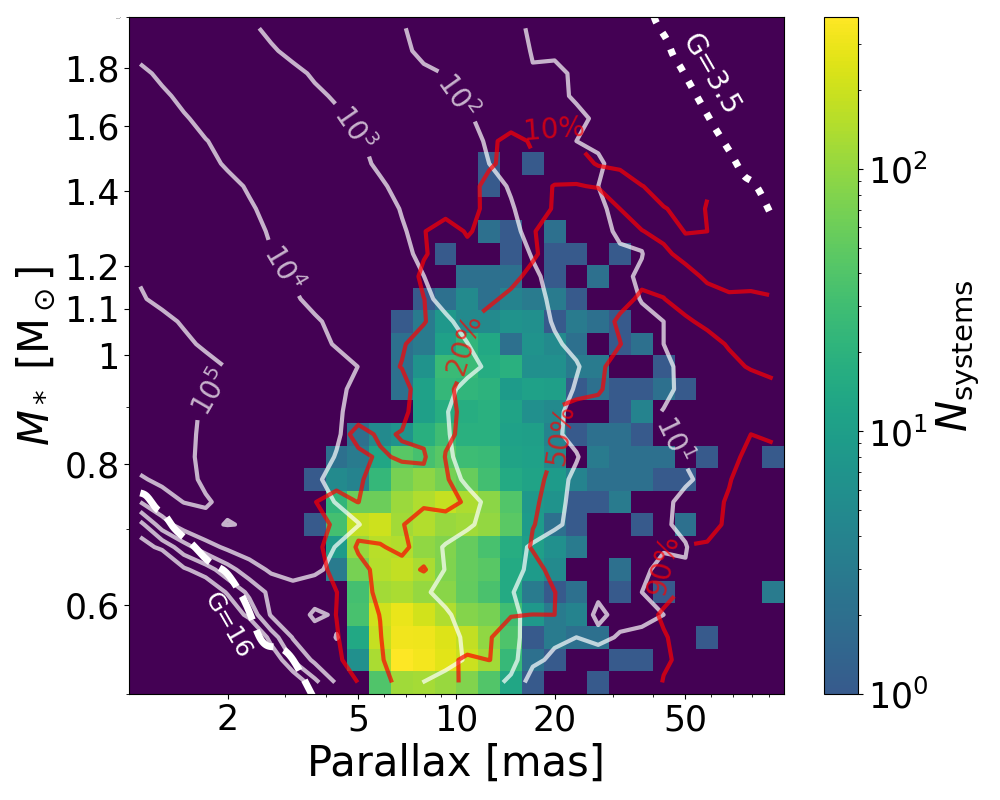}
    \caption{Number of detected planet candidates (per bin) with respect to parallax and $M_\star$. The sensitivity curve of \textit{Gaia} for detecting 2--13.5 M$_{\rm J}$ companions at 1--3\,au beyond a significance of 2.7--$\sigma$ is overplotted in red showing the regions where more than 10, 20, 50 and 90\% of companions are detected. The distribution of the \textit{Gaia} sources satisfying steps 1--3 (Fig.~\ref{fig:steps}) is also overplotted in white solid lines. The dashed thick white lines bound the region within the magnitude range of our sample, that is $G$-mag within 3.5--16.}
    \label{fig:plx_mstar}
\end{figure}

Figure~\ref{fig:plx_mstar} represents the number of planet-candidate hosts with respect to $\varpi$ and $M_{\star}$. It can be seen that all planet candidate hosts   are indeed close-by with $\varpi$$>$2.9\,mas ($d$$<$345\,pc), and that most planet candidates are found around stars with $M_{\star}$$<$0.8\,M$_\odot$ and $\varpi$$<$20\,mas; that is, M and K-dwarfs beyond 50\,pc. This is the result of both the increase in the volume of stars with increasing distance and the increase in the detectability of planets around fainter stars with smaller $M_{\star}$ and larger $\varpi$. This is also clear when comparing the distribution of planet-candidate hosts to the distribution of sources in the input sample at step 3 (shown as white contours), and the modulation of the sensitivity of \textit{Gaia}'s \glsxtrshort{ruwe} for the detection of 2-13.5 M$_{\rm J}$ companions within 1--3\,au (shown as red contours) with respect to $M_{\star}$ and $\varpi$. The sensitivity levels are obtained as explained in Sect. 9 of Paper I. On a grid of 30$\times$30 bins of $M_{\star}$ and $\varpi$, we simulated ---for each bin--- 1000 values of \glsxtrshort{mse} for our reference system GJ\,832 assuming a companion with a mass of within 2--13.5\,M$_{\rm J}$ and an \glsxtrshort{sma} of within 1--3\,au. The sensitivity, at any $M_{\star}$ and $\varpi$, is the percentage of simulations with an \glsxtrshort{mse} more significant than 2.7--$\sigma$.  

\begin{figure}[hbt]
    \centering
    \includegraphics[width=89.3mm]{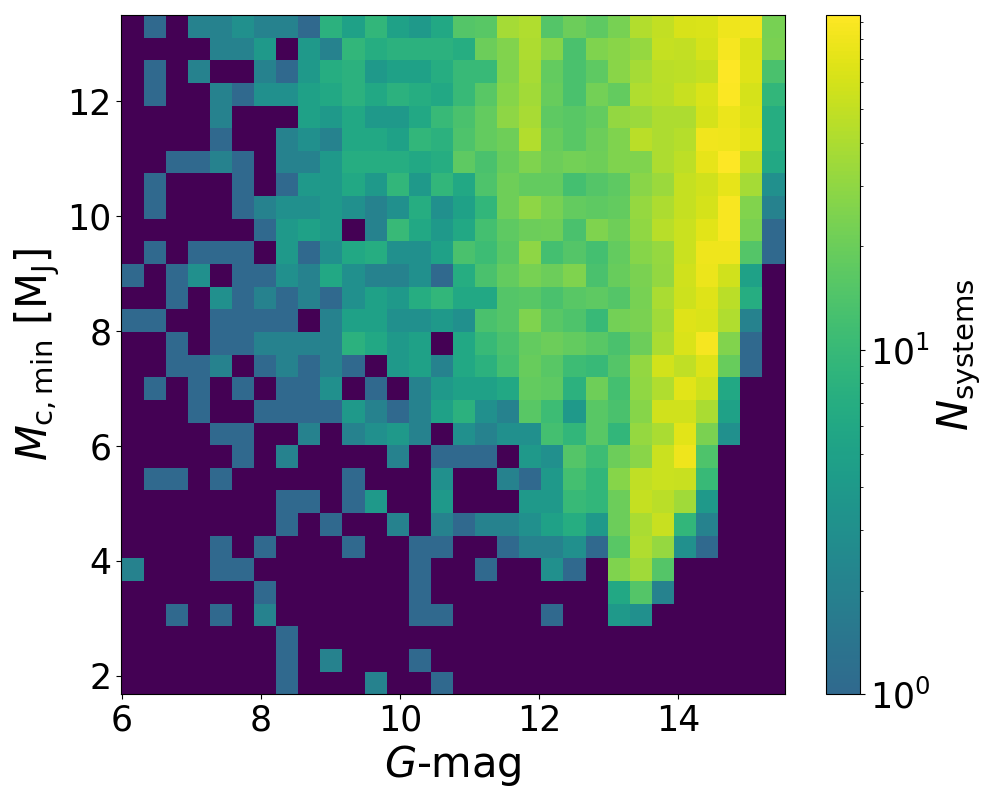}
    \includegraphics[width=89.3mm]{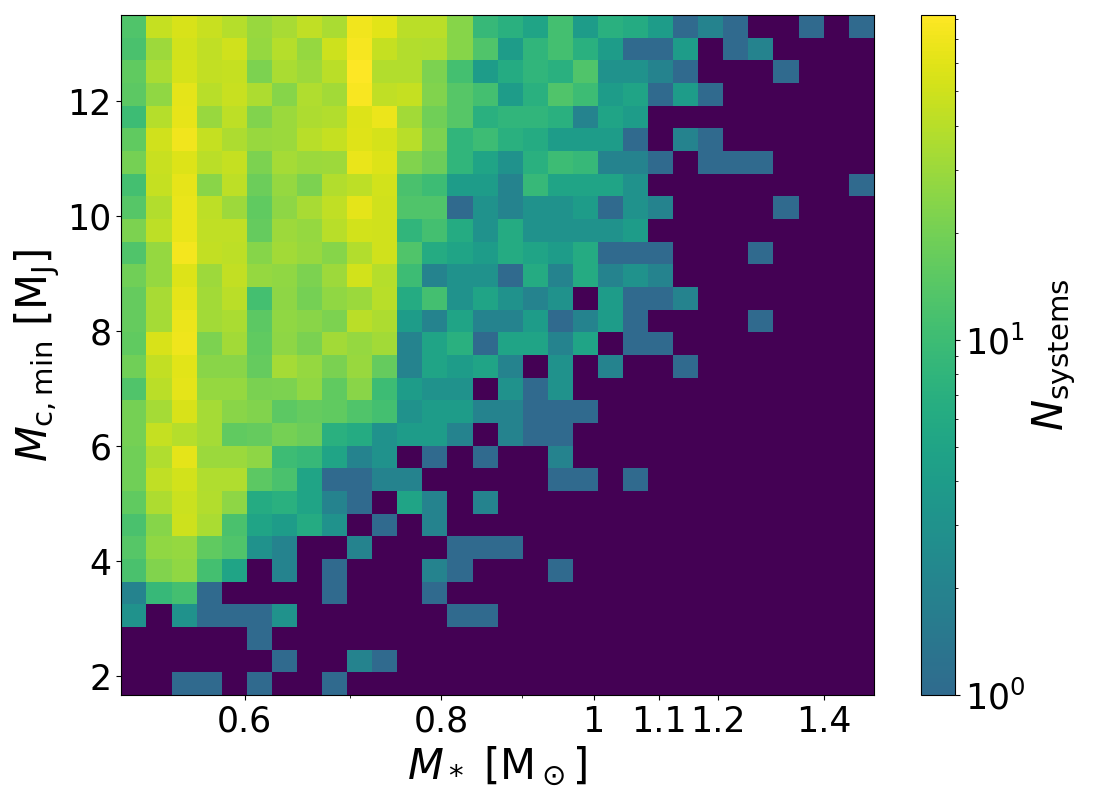}
    \caption{Number of detected planet-candidate hosts (per bin) with respect to the minimum mass of the  companion and the $G$-magnitude or mass of the host star (top and bottom, respectively). }
    \label{fig:mag_mpmin_planets_densityplot}
\end{figure}

Figure~\ref{fig:mag_mpmin_planets_densityplot} shows the distribution of the minimum mass of the planet candidates with respect to the $G$-mag and $M_\star$ of the hosts. Planet candidates with $M_{\rm c,min}<5$\,M$_{\rm J}$ are mostly detected around sources with $G$$\sim$14. Those correspond to low-mass stars, namely MK-dwarfs with $M_\star$$<$0.7\,M$_\odot$. We did not include M-dwarfs with masses of $<$0.5\,M$_\odot$ because we used the \verb+mass-Flame+ of the CU8 catalog, which is only determined for stars with masses of $>$0.5\,M$_\odot$. This explains why we do not detect planet candidates with $M_{\rm c,min}<5$\,M$_{\rm J}$ beyond $G$=14. Nonetheless, M-dwarfs offer the best opportunity to detect exoplanets with \textit{Gaia}. 

\begin{figure}[hbt]
    \centering
    \includegraphics[width=89.3mm]{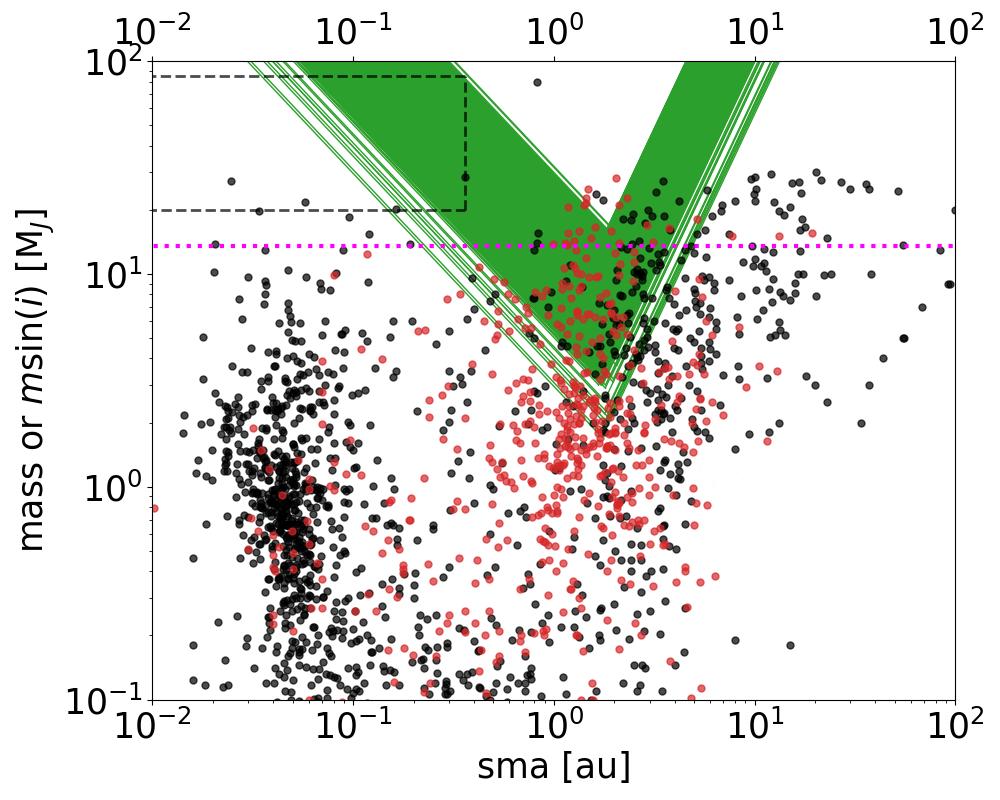}
    \caption{Mass--sma diagram of known exoplanets and planet candidates presented in the current work. The green lines show the degenerate mass and sma compatible with the \glsxtrshort{g3} $\alpha_{\rm \glsxtrshort{mse}}$. The data for  exoplanets are taken from the \glsxtrshort{nea}, including those with a known mass (black dots) and those with only a minimum mass $m\sin i$ (red dots). The black dashed lines bound the BD desert region extending up to orbital periods of  80 days ($\sim$0.36\,au for a 1\,M$_\odot$ host star;~\citealt{Kiefer2019a,Kiefer2021}). The magenta dotted line represents the commonly accepted 13.5 M$_{\rm J}$ upper-mass limit on the planetary domain.}
    \label{fig:mass_sma_diagram}
\end{figure}

The candidate exoplanets are compared to the known exoplanet population taken from the NASA Exoplanet Catalog\footnote{\url{https://exoplanetarchive.ipac.caltech.edu}} in the mass--sma diagram presented in Fig.~\ref{fig:mass_sma_diagram}. Because of the geometry of the V-shaped curve, whose minimum lies at about 2\,au, the sma of the exoplanet candidates are all contained at most within 0.1--10\,au. Extrapolating along the empirical mass--sma relationships presented in Sect.~\ref{sec:pmex}, we delineated the wider region of possible mass and sma spanned by the exoplanet candidates with sma$\neq$2\,au. This exercise shows that the current domain of sensitivity of \textit{Gaia} allows us to probe exoplanets that can be further confirmed and characterized using RVs. Moreover, the set of solutions covers a domain of orbits for Jupiter-like exoplanets that are less represented between 0.1 and 1\,au, which is on the border of the orbit circularization zone~\citep{Kane2013}. The full set of solutions extend to within the BD desert below orbital periods of  80 days~\citep{Kiefer2019a,Kiefer2021}, that is sma$<$0.36\,au for a 1M$_\odot$ host star. If some of the candidates are found to not be exoplanets and with an sma$<$0.4\,au, then they may become newly identified BD.

\section{Discussion}
\label{sec:discussion}

The complete vetting of our catalog of 9,698 planet-candidate hosts will be the subject of future studies. Nevertheless, we searched the \glsxtrshort{nss}~\citep{Arenou2023}, and found 339 sources with an orbital solution among our 9,698 planet-candidate hosts. Moreover, we searched the \glsxtrlong{cds} (\glsxtrshort{cds}) and found 2,951 sources referenced in Simbad, including 286 that were published in the Washington Double Star catalog (WDS;~\citealt{Mason2001}). These latter sources are listed in Table~\ref{tab:planet_catalog}, where the \glsxtrshort{nss} sources are flagged and the WDS names are given when known. In Sect.~\ref{sec:holl2023} we describe the \glsxtrshort{ruwe}-based astrometric signatures and the astrometric orbital solutions of the validated \glsxtrshort{nss} orbits presented in~\citet{Holl2023}, and cross-check our catalog of planet candidate hosts with their estimation of the companion mass. In Sect.~\ref{sec:xmatch_nea} we cross-match our catalog with the exoplanets listed in the \glsxtrshort{nea}. Finally, in Sect.~\ref{sec:xmatch_HIP} we focus on those of our selected systems that were observed with \textsc{Hipparcos} and for which the \glsxtrshort{pma} can help in characterizing the sma and mass of the identified companion. 

\subsection{A comparison with \textit{Gaia}-NSS-validated orbits and planet-candidate cross-match}
\label{sec:holl2023}

Holl et al. (2023; H23) identified and validated the astrometric orbits for 204 sources published in the \glsxtrshort{nss} catalog with an orbital period of $<$5.6\,yr, and further characterized in~\citet{Arenou2023}. We compared the \glsxtrshort{ruwe}-based astrometric signatures determined in this work to those deduced from the astrometric orbits, and cross-checked our planet selection with the classification of H23 according to the estimated companion mass.

\subsubsection{Comparing the \texorpdfstring{$\alpha_{\rm \glsxtrshort{mse},\glsxtrshort{ruwe}}$}{\texttt{ruwe}-based astrometric signature} and the orbital solutions}

Among the 204 sources of H23, two have a $G$-mag of $>$16 and hence are not part of our present sample, namely DENIS J082303.1-491201 and 2MASS J08053189+4812330, two BDs with respective $G$-mag of 18.5 and 20. The other 202 sources have $G$$<$16, and are listed in Table~D.1. Among them, 196 are well behaved, verifying criteria $C_0 \cap C_1$, as defined in Fig.~\ref{fig:steps}, 101 of which have a \verb+mass-Flame+ characterized in the CU8 database and are located on the main sequence (criteria $C_0 \cap  C_1 \cap C_2$).  We find that $\sim$93\% of the H23 sample, that is 184 of the 196 well-behaved sources, including 93 of the 101 with a \verb+mass-Flame+, have an $\alpha_{\rm \glsxtrshort{mse},\glsxtrshort{ruwe}}$ that is more significant than 2.7--$\sigma$ (criteria $C_0 \cap C_1 \cap C_4$). Only 12 sources are found to meet all the selection criteria of our planet-candidate hosts sample.

H23 classified the companions of their 204 sources into three categories with respect to their mass range if their host star had a mass of 1\,M$_\odot$: pseudo-mass-index=0 if $<$20\,M$_{\rm J}$ (11 sources), pseudo-mass-index=1 if within 20-120\,M$_{\rm J}$ (32 sources), and pseudo-mass-index=2 if $>$120\,M$_{\rm J}$ (161 sources). At a pseudo-mass-index of 0, of the 10 well-behaved sources, 5 have an
$\alpha_{\rm \glsxtrshort{mse},\glsxtrshort{ruwe}}$ more significant than
2.7--$\sigma$. At a pseudo-mass-index of 1, that is 29 among 31 well-behaved sources. At a pseudo-mass-index of 3, that is 150 among 156 well-behaved sources. Thus, in 94-96\% of the cases, companions in the BD and stellar mass regime that are detected and characterized with the \textit{Gaia} astrometric time series are associated with an $\alpha_{\rm \glsxtrshort{mse},\glsxtrshort{ruwe}}$ more significant than 2.7--$\sigma$. 
In the planetary regime, at a pseudo-mass index of 0, this rate is $\sim$50\%. This means that, unsurprisingly, due to the necessity of defining a threshold on the significance of $\alpha_{\rm \glsxtrshort{mse},\glsxtrshort{ruwe}}$, a large fraction of planetary companions were missed. 

Within the H23 sample, we find 11 sources with an $\alpha_{\rm \glsxtrshort{mse},\glsxtrshort{ruwe}}$ that cannot reject the single-star hypothesis, at a $p$-value of smaller than the 2.7--$\sigma$ threshold. More specifically, we identify:
\begin{itemize}
    \item AK\,For, HD\,106770, HD\,188622, and HD\,211419, which have an $\alpha_{\rm \glsxtrshort{mse}, \glsxtrshort{ruwe}}$ significance of within 2--2.7--$\sigma$; whether they can be considered as detected or not is disputable, because the 2--$\sigma$ threshold could also have been adopted, more conventionally, so as to reject the null hypothesis with a $p$-value of $<$4.6\%. We recall that, for the sake of keeping an \glsxtrshort{fp} rate below 10\%, we had to adopt a more restrictive threshold of 2.7--$\sigma$;
    \item HD\,142, $\iota$\,Hor, TYC\,8841-182-1, HIP\,66074 (\textit{Gaia}-3), HD\,184962, HD\,100069, and HD\,132406, each with a significance smaller than 2--$\sigma$; 
    \item  HD\,142 and AK\,For, for which the \glsxtrshort{pma} rejects the single-star hypothesis with a significance of respectively 2.8--$\sigma$ and more than 9--$\sigma$: \glsxtrshort{pmex} thus detects the known companion in both cases.
\end{itemize}

\begin{figure}
    \centering
    \includegraphics[width=89.3mm,clip=true]{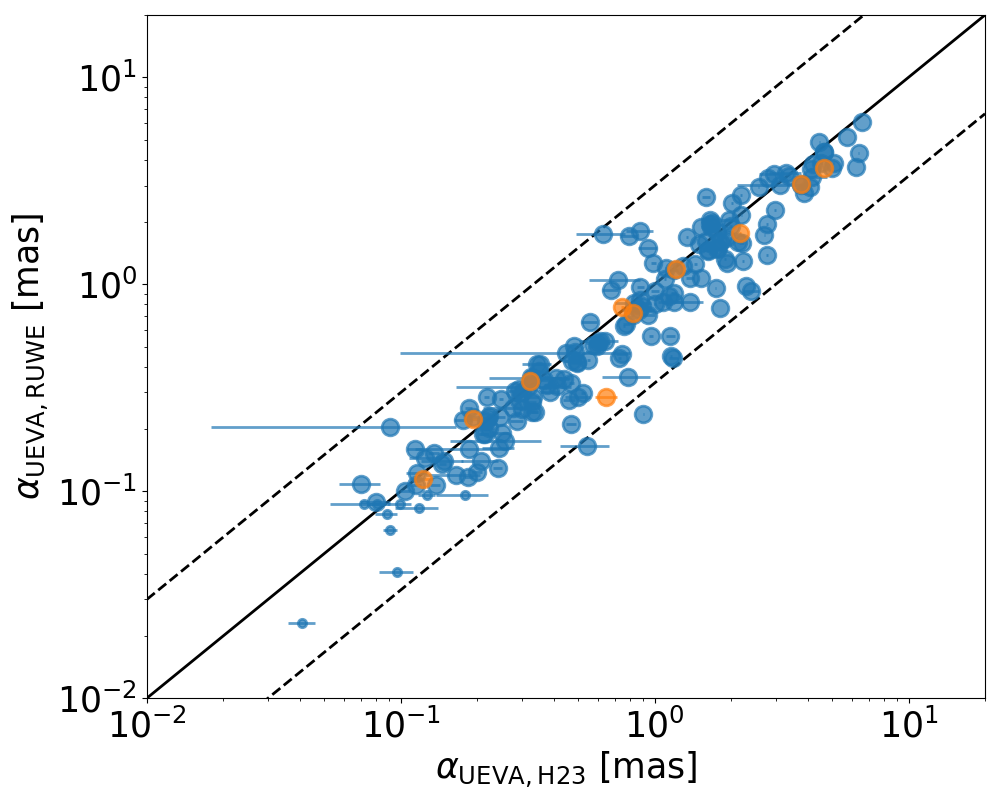}
    \caption{Comparing the $\alpha_{\rm \glsxtrshort{mse},\glsxtrshort{ruwe}}$ determined in the \glsxtrshort{g3} to those calculated from the known astrometric parameters of a companion in~\citet{Holl2023} and~\citet{Arenou2023}. The blue and orange circles show the sources from respectively the 5p and 6p datasets. Tiny-sized symbols indicate a significance of $\alpha_{\rm \glsxtrshort{mse},\glsxtrshort{ruwe}}$$<$2.7--$\sigma$. The solid line shows the equality, and the dashed lines depict factors of 3 differences, between the data of the two axis.}
    \label{fig:compare_holl2023}
\end{figure}

\noindent The scale of the $\alpha_{\rm \glsxtrshort{mse}}$ expected from the mass and period of their companions and the parallax can explain this lower significance. Figure~\ref{fig:compare_holl2023} compares the $\alpha_{\rm \glsxtrshort{mse},\glsxtrshort{ruwe}}$ to the $\alpha_{\rm \glsxtrshort{mse},H23}$ calculated from the known \textit{Gaia} astrometric solutions of the H23 list in~\citet{Arenou2023} and the empirical relations in Eq.~\ref{eq:mass_sma_AEN}, as explained in Appendix~\ref{sec:details_holl2023}. 
The $\alpha_{\rm \glsxtrshort{mse},\glsxtrshort{ruwe}}$ and the $\alpha_{\rm \glsxtrshort{mse},H23}$ generally agree to within an order of magnitude in the range 0.1--10\,mas. Therefore, our estimation of the astrometric signatures from the RUWE is generally compatible with the mass and sma of companions around sources with fitted astrometric orbits.  

Apparent counter-examples to this statement are HD\,134251, HD\,108510, HD\,221757, HD\,117126, and HD\,8054, which all lead to overestimation of $\alpha_{\rm \glsxtrshort{mse},H23}$
compared to $\alpha_{\rm \glsxtrshort{mse},\glsxtrshort{ruwe}}$by  a factor of $>$2.5 and even $>$3 (HD\,134251 and HD\,108510). These are all binaries (pseudo-mass index of 2). We find no satisfactory explanation for this discrepancy, while \glsxtrshort{rv}-SB1 solutions tend to validate the orbital parameter and mass of the companion used to calculate $\alpha_{\rm \glsxtrshort{mse},H23}$.

\begin{figure}
    \centering
    \includegraphics[width=89.3mm,clip=true]{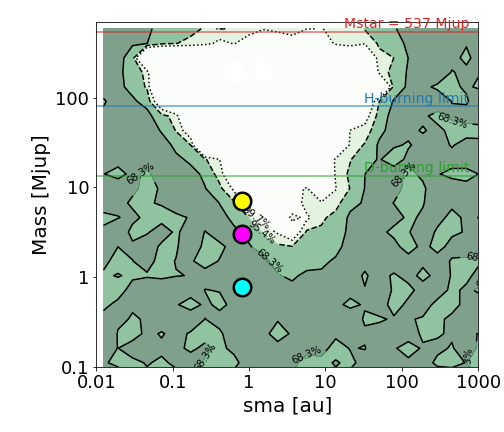}
    \caption{Same as Fig.~\ref{fig:PMEX_HD40503} but for HIP\,66074. Individual maps based on either \glsxtrshort{pma} or \glsxtrshort{ruwe} only are shown in Fig.~E.2. The yellow and cyan points show respectively the 7 and 0.79\,M$_{\rm J}$ solutions at 0.8\,au from the \glsxtrshort{nss} and the \glsxtrshort{rv}. The magenta point shows the solution at 3\,M$_{\rm J}$ if the photocentric sma measured in the \glsxtrshort{nss} is overestimated by a factor 2, corresponding to an inclination of 13$^{\circ}$.}
    \label{fig:PMEX_HIP66074}
\end{figure}

In that regard, the case of HIP\,66074, which has $\alpha_{\rm \glsxtrshort{mse},H23}$/$\alpha_{\rm \glsxtrshort{mse},\glsxtrshort{ruwe}}$=2.4, is interesting, as its companion is \textit{Gaia}-3\,b, a planet~\citep{Winn2022,Marcussen2023}. Here, the mismatch might be related to an issue with the astrometric solution published in the \glsxtrshort{nss}, which is inconsistent with the \glsxtrshort{rv} solutions for this system. Both predict an edge-on inclination, but do not agree on the mass of the companion \textit{Gaia}-3\,b, suggesting 7.3\,M$_{\rm J}$ (\glsxtrshort{nss}) and 0.4\,M$_{\rm J}$ (\glsxtrshort{rv}). From the \glsxtrshort{nss} solution, we predict $\alpha_{\rm \glsxtrshort{mse},H23}$=0.093$\pm$0.014\,mas, while we find a 1.1--$\sigma$ significant  $\alpha_{\rm \glsxtrshort{mse},\glsxtrshort{ruwe}}$=0.041\,mas. Such a small insignificant $\alpha_{\rm \glsxtrshort{mse}}$ would rather be compatible with the small-mass estimated by \glsxtrshort{rv}. This is in tension with the results from~\citet{Sozzetti2023}, who propose, to reconcile the RV and \glsxtrshort{nss} solutions, a companion with a mass of 3--7\,M$_{\rm J}$ on a face-on orbit. This solution would still imply a much larger value of $\alpha_{\rm \glsxtrshort{mse}}$  of close to 0.1\,mas, with a significance certainly beyond 2.7--$\sigma$. Nevertheless, \citet{Sozzetti2023} proposed that the sma of the photocenter in the \glsxtrshort{nss} solution could be overestimated by up to a factor $\sim$2. In such a case, the astrometric signatures would indeed better match, with now $\alpha_{\rm \glsxtrshort{mse},H23}$=0.047$\pm$0.007\,mas. According to \citet{Sozzetti2023}, this would suggest an inclination of $\sim$13$^\circ$ for this planet, and a mass of $\sim$3\,M$_{\rm J}$. Figure~\ref{fig:PMEX_HIP66074} shows the \glsxtrshort{pmex} map obtained for HIP\,66074, combining \glsxtrshort{ruwe} and \glsxtrshort{pma} and comparing these to the possible masses of the companion \textit{Gaia}-3\,b.

\subsubsection{A cross-match of our planet catalog with the H23 classification}

 We cross-matched the H23 sample with our catalog of planet-candidate hosts. Among the eight sources at a pseudo-mass index of zero and meeting the criteria $C_0\cap C_1\cap C_2$, we find five that belong to our catalog. Moreover, we find that 7 of the 93 sources with a pseudo-mass index 1 or 2  also belong to our catalog. As expected, because of the mass--sma degeneracy, our selection process indeed picks up systems with companions that have a mass above the planetary regime. 
The five sources listed in our catalog at a pseudo-mass index of zero  are HD\,40503 (with ID \# 212), HD\,164604 (\# 302), HD\,111232 (\# 42), HD\,81040 (\# 56), and HD\,175167 (\# 74). Four companions, HD\,164604\,b, HD\,111232\,b, HD\,81040\,b and HD\,175167\,b, are already known exoplanets and are discussed in Sects.~\ref{sec:xmatch_nea} and~\ref{sec:xmatch_HIP}. 

The criteria $C_4$ and $C_5$ applied on the H23 input sample of 101 systems with a characterized CU8 \texttt{mass-Flame} led to the selection of 12 systems, that is 12\% of this input sample. Among the \glsxtrshort{bd}/star companions,  criteria $C_4$ and $C_5$  led to the selection of  7.5\% of them, and among the planets, the same criteria led to the selection of 62.5\% of them. This implies that our criteria for selecting systems with a possible planetary mass companion within a sample of systems whose $M_\star$ is known are about eight times more likely to select planets than to select \glsxtrshort{bd} or stellar companions in the range of orbits where \textit{Gaia} is sensitive. 
This shows that the significance of $\alpha_{\rm \glsxtrshort{mse}}$ is a valid criterion for identifying companions whose orbits can be validated afterward, either using astrometric time series published with the \glsxtrlong{dr4} (\glsxtrshort{dr4}) or other means such as RV or direct imaging. 

We conclude that, even though we have identified a few apparent inconsistencies between the observed astrometric signature and the orbital fit of the astrometric time series, they generally agree. We can thus expect that the orbital astrometric motion of most of the planet-candidate-host stars identified in Sect.~\ref{sec:catalog} that have $\alpha_{\rm \glsxtrshort{mse},\glsxtrshort{ruwe}}$$>$0.1\,mas and a period of $<$10\,yr could be characterized with the publication of the astrometric time series with the \glsxtrshort{dr4}.

\subsection{A cross-match with planet and binary catalogs}
\label{sec:xmatch_nea}

Among the 5,678 planets from 4,236 planetary systems published in the \glsxtrshort{nea}, we find 3,793   sources  to be brighter than $G$=16 in the \glsxtrshort{5p} and \glsxtrshort{6p} datasets of the \glsxtrshort{g3} catalog. Only 2,423 have a \verb+mass-Flame+, and among those, 149 have an $\alpha_{\rm \glsxtrshort{mse},\glsxtrshort{ruwe}}$ more significant than 2.7--$\sigma$. Finally, 19 systems have an astrometric minimum mass of below 13.5\,M$_{\rm J}$ and belong to our sample of planet-candidate hosts, all of them in the \glsxtrshort{5p} dataset. We note that therefore 130\,NEA planetary systems must host a BD or stellar companion, instead or on top of the existing planetary companion(s) in those 130 systems. The detailed study of this binary subsample will be within the scope of a future study. We list the 19 planetary systems present in our 9,698 sample in Table~\ref{tab:planet_catalog}, and give details on each below. We determined the \glsxtrshort{pmex} maps from the \glsxtrshort{ruwe} in Fig.~F.1. For the sources that were also observed by \textsc{Hipparcos}, the constraints from \glsxtrshort{pma} could also be used in combination with the \glsxtrshort{ruwe}. 

\begin{itemize}[align = left]
    \item [\it 2MASS J04372171+2651014 (\# 6883).] This low-mass pre-main sequence 2.5$\pm$0.4-Myr-old M-dwarf located at 128\,pc is known to host a 4 M$_{\rm J}$ companion, 2M0437\,b, at $\sim$118$\pm$1.3\,au~\citep{Gaidos2022}. The \glsxtrshort{ruwe}=1.21 for this 14.3 mag star corresponds to an $\alpha_{\rm \glsxtrshort{mse}, \glsxtrshort{ruwe}}$=0.096\,mas with a 3.3--$\sigma$ significance. The \glsxtrshort{pmex} map for this star seems to report the detection of a companion at less than 100\,pc. However, we would tend to be rather cautious in this case as, for a  star as  young as 2M0437, one may expect significant accretion, which could induce an astrometric jitter if $L_{\rm acc}/L_\star$$>$10$^{-3}$. We also noted in the epoch photometry available from the online single-object search engine of the \textit{Gaia} archives that two photometric points in the $B$ and $R$ bands (on 18 Oct 2015 and 5 Aug 2016) were anomalous. This may indicate possible issues with the corresponding astrometric points too.\\
    
    \item [\it HD 111232 (\# 42).] This G8V star located at 29\,pc is known to host two \glsxtrshort{rv} companions~\citep{Mayor2004}, including one planet, HD 111232\,b, of 7.965$^{+1.128}_{-0.479}$\,M$_{\rm J}$ at 2.148$^{+0.088}_{-0.097}$\,au, and one brown dwarf, of 18.1$^{+4.2}_{-1.6}$\,M$_{\rm J}$ at 17.25$^{+2.158}_{-2.151}$\,au~\citep{Feng2022}. Our \glsxtrshort{pmex} map for HD\,111232, combining the constraints from  \glsxtrshort{ruwe} (1.24; 3.6--$\sigma$) and \glsxtrshort{pma} (0.54$\pm$0.03\,mas\,yr$^{-1}$; $>$9--$\sigma$), predicts a 4--50\,M$_{\rm J}$ companion within 2--10\,au of the star with 95.4\% confidence. This prediction is in nice agreement with the published parameters of HD 111232\,b. \\
    
    \item [\it HD 136118 (\# 16).] This F9V star located at 52\,pc is known to host one \glsxtrshort{rv} companion of $M\sin i$$\sim$11.9\,M$_{\rm J}$ and \glsxtrshort{sma}$\sim$2.3\,au~\citep{Fischer2002}. Using the Hubble Fine Guidance Sensor,~\citet{Martioli2010} further showed that this planet candidate is a true BD with a mass of 42$^{+11}_{-8}$\,M$_{\rm J}$ at 2.36$\pm$0.05\,au. Our \glsxtrshort{pmex} map for HD\,136118, combining the constraints from \glsxtrshort{ruwe} (1.43; 3.3--$\sigma$) and \glsxtrshort{pma} (0.43$\pm$0.03\,mas\,yr$^{-1}$; 5.1--$\sigma$), predicts a $>$8--M$_{\rm J}$ companion with an \glsxtrshort{sma}$<$10\,au with 95.4\% confidence. It moreover predicts a mass of within 8--40\,M$_{\rm J}$ if the sma$\sim$2.4\,au. This is in better agreement with the published parameters for HD\,136118\,b within error bars than the findings of~\citet{Feng2022}, who predicted a mass within 13.10$^{+1.35}_{-1.27}$\,M$_{\rm J}$. \\
    
    \item [\it HD 13808 (\# 89).] This K2V star located at 29\,pc is known to host two \glsxtrshort{rv} planets~\citep{Mayor2011,Ahrer2021} with minimum masses of 0.03599$\pm$0.0025\,M$_{\rm J}$ at 0.11\,au, and 0.0315$\pm$0.0038\,M$_{\rm J}$ at 0.26\,au. Our \glsxtrshort{pmex} map for HD\,13808, combining the constraints from \glsxtrshort{ruwe} (1.17; 2.9--$\sigma$) and \glsxtrshort{pma} (0.037$\pm$0.026\,mas\,yr$^{-1}$; 0.4--$\sigma$), predicts a $>$2--M$_{\rm J}$ companion within 2\,au of the star with 95.4\% confidence. It moreover predicts a mass rather in the BD domain in this 2--$\sigma$ confidence region, for at least one of the two planet candidates. This would imply an almost face-on system with an inclination of $<$0.3$^{\circ}$. With an insignificant \glsxtrshort{pma}, this result is mainly driven by the \glsxtrshort{ruwe}, which corresponds to $\alpha_{\rm \glsxtrshort{mse},\glsxtrshort{ruwe}}$=0.12\,mas. \\
    
    \item [\it HD 164604 (\# 302).] This K3.5V star located at 39\,pc is known to host one \glsxtrshort{rv} planet~\citep{Arriagada2010}, with $M\sin i$=2.7$\pm$1.3\,M$_{\rm J}$ at 1.3$\pm$0.5\,au. The orbit directly fitted to the astrometric points in the \glsxtrshort{g3}~\citep{Arenou2023} led to a larger mass of 14$\pm$5.5\,M$_{\rm J}$. Our \glsxtrshort{pmex} map for HD\,164604, combining the constraints from \glsxtrshort{ruwe} (1.16; 6.3--$\sigma$) and \glsxtrshort{pma} (0.55$\pm$0.08\,mas\,yr$^{-1}$; 6.2--$\sigma$), predicts a $>$5--M$_{\rm J}$ companion with an sma of $<$10\,au with 95.4\% confidence. At 1.3\,au, \glsxtrshort{pmex} predicts a mass of within 10--30\,M$_{\rm J}$, in agreement with the results of~\citet{Arenou2023}.  \\
    
    \item [\it HD 175167 (\# 74).] This G5IV/V star located at 71\,pc is known to host one \glsxtrshort{rv} planet~\citep{Arriagada2010}, with $M\sin i$=7.8$\pm$3.5\,M$_{\rm J}$ at 2.4$\pm$0.05\,au. The orbit directly fitted to the astrometric points in the \glsxtrshort{g3}~\citep{Arenou2023,Winn2022} led to a larger mass of 14.8$\pm$1.8\,M$_{\rm J}$. Later combination with MIKE+FPS by~\citet{Gan2023} led to a slightly lower mass of 10.2$\pm$0.4\,M$_{\rm J}$. Our \glsxtrshort{pmex} map for HD\,175167 combining the constraints from \glsxtrshort{ruwe} (1.17; 3.1--$\sigma$) and \glsxtrshort{pma} (0.19$\pm$0.02\,mas\,yr$^{-1}$; 4.1--$\sigma$) predicts a $>$5--M$_{\rm J}$ companion with an sma mostly $<$10\,au with 95.4\% confidence. At 2.4\,au, \glsxtrshort{pmex} predicts a mass within 6--20\,M$_{\rm J}$ with 95.4\% confidence and 7--15\,M$_{\rm J}$ with 68.3\% confidence. This agrees well with the results from~\citet{Gan2023}.  \\

    \item [\it HD 221287 (\# 64).] This F7V star located at 53\,pc is known to host one \glsxtrshort{rv} companion of $M\sin i$=3.1$\pm$0.8\,M$_{\rm J}$ and \glsxtrshort{sma}=1.25$\pm$0.4\,au~\citep{Naef2007}. To our knowledge, this \glsxtrshort{rv} planet has never been confirmed. Our \glsxtrshort{pmex} map for HD\,136118, combining the constraints from \glsxtrshort{ruwe} (1.2; 3--$\sigma$) and \glsxtrshort{pma} (0.026$\pm$0.027\,mas\,yr$^{-1}$; 0.34--$\sigma$), predicts a $>$3--M$_{\rm J}$ companion with an \glsxtrshort{sma} of $<$3\,au with 95.4\% confidence. It moreover predicts a mass of within 3--40\,M$_{\rm J}$ if the sma is $\sim$1.3\,au at 95.4\% confidence and within 10-20\,M$_{\rm J}$ at 68.3\% confidence. This agrees with the published \glsxtrshort{rv}-derived parameters for HD\,221287\,b but does not confirm the planetary nature of this object. \\
    
    \item [\it HD 23596 (\# 29).] As already discussed in Paper I, for this 7.2-mag F8 star at 52\,pc, the combination of \glsxtrshort{pma} (0.59$\pm$0.04\,mas\,yr$^{-1}$; 7.1--$\sigma$) and \glsxtrshort{ruwe} (1.35; 3.5--$\sigma$) led \glsxtrshort{pmex} to infer a companion in the BD domain, with a narrow constraint on mass of  10--30\,M$_{\rm J,}$ as well as on \glsxtrshort{sma}, namely 2--5\,au at 68.3\% confidence. This is in perfect agreement with the known companion of HD\,23596 at 2.90$\pm$0.08\,au, first discovered as an 8.2 M$_{\rm J}$ super-Jupiter with the ELODIE spectrograph~\citep{Perrier2003}, and further re-established as a 14 M$_{\rm J}$ low-mass BD combining RVs and \textsc{Hipparcos}--\textit{Gaia} \glsxtrshort{pma}~\citep{Feng2022,Xiao2023}. \\
    
    \item [\it HD 28254 (\# 51).] This G1IV/V star located at 56\,pc is known to host one \glsxtrshort{rv} planet~\citep{Naef2010}, with $M\sin i$=1.16$^{+0.1}_{-0.06}$\,M$_{\rm J}$ at 2.15$\pm$0.05\,au. Combining \glsxtrshort{rv} with \textsc{Hipparcos}--\textit{Gaia} astrometry led to a larger mass of 1.5--6.5\,M$_{\rm J}$~\citep{Philipot2023b}. Our \glsxtrshort{pmex} map for HD\,28254, combining the constraints from \glsxtrshort{ruwe} (1.51; 5.2--$\sigma$) and \glsxtrshort{pma} (0.18$\pm$0.04\,mas\,yr$^{-1}$; 2.4--$\sigma$), predicts a $>$6--M$_{\rm J}$ companion with an sma of $<$4\,au with 95.4\% confidence. At 2.15\,au, \glsxtrshort{pmex} predicts a mass of within 6--30\,M$_{\rm J}$ with 95.4\% confidence and 9--15\,M$_{\rm J}$ with 68.3\% confidence. This is only marginally compatible with the results of~\citet{Philipot2023b}, but their posterior distribution on the companion mass had a long tail toward larger mass. This indicates that the constraint from \glsxtrshort{ruwe} leads to an even higher mass for HD\,28254\,b.    \\
    
    \item [\it HD 62364 (\# 30).] This F7V star located at 53\,pc is known to host one \glsxtrshort{rv} companion of $M\sin i$=12.7$\pm$0.2\,M$_{\rm J}$ and \glsxtrshort{sma}=6.15$\pm$0.04\,au but constrained from \textsc{Hipparcos}--\textit{Gaia} astrometry to a higher BD mass of 18.77$\pm$0.66\,M$_{\rm J}$~\citep{Frensch2023}. Our \glsxtrshort{pmex} map for HD\,62364, combining the constraints from \glsxtrshort{ruwe} (1.46; 5--$\sigma$) and \glsxtrshort{pma} (0.79$\pm$0.03\,mas\,yr$^{-1}$; $>$9--$\sigma$), predicts a 15--200\,M$_{\rm J}$ companion with an \glsxtrshort{sma} of within 2--20\,au with 95.4\% confidence. This agrees at 2--$\sigma$ with the published parameters for HD\,62364\,b. \\

    \item [\it HD 81040 (\# 56).] As already discussed in Paper I, for this 7.2 mag G2/3V star at 33\,pc, the combination of \glsxtrshort{pma} (0.15$\pm$0.05\,mas\,yr$^{-1}$; 1.5--$\sigma$) and \glsxtrshort{ruwe} (1.60; 6.8--$\sigma$) led \glsxtrshort{pmex} to infer a companion with a mass of possibly as low as 6\,M$_{\rm J}$ with an \glsxtrshort{sma} smaller than 4\,au at 95.4\% confidence. This was in good agreement with the known companion of HD\,81040 at 1.94\,au, first discovered as an 6.9 M$_{\rm J}$ super-Jupiter by~\citet{Sozzetti2006}, and further confirmed at a mass of 8.04$_{-0.54}^{+0.66}$\,M$_{\rm J}$ combining RVs and \textit{Gaia} orbit fit of the astrometric time series~\citep{Arenou2023}\footnote{See also the dedicated webpage on the \textit{Gaia} ESA website at \url{https://www.cosmos.esa.int/web/gaia/iow\_20220131}}. \\
    
    \item [\it HD 9446 (\# 100).] This G5V star located at 53\,pc is known to host two \glsxtrshort{rv} planets~\citep{Hebrard2010}, with 0.7$\pm$0.06\,M$_{\rm J}$ at 0.189$\pm$0.006\,au, and 1.82$\pm$0.17\,M$_{\rm J}$ at 0.654$\pm$0.022\,au. Our \glsxtrshort{pmex} map for HD\,9446, combining the constraints from \glsxtrshort{ruwe} (1.22; 2.8--$\sigma$) and \glsxtrshort{pma} (0.11$\pm$0.05\,mas\,yr$^{-1}$; 1.1--$\sigma$), predicts a $>$1.5--M$_{\rm J}$ companion within 7\,au from the star with 95.4\% confidence. It moreover predicts a mass of greater than 8\,M$_{\rm J}$ in this 2--$\sigma$ confidence region, for at least one of the two planet candidates. This would imply a system close to face on, with an inclination of $<$15$^\circ$. With an insignificant \glsxtrshort{pma}, this result is mainly driven by the \glsxtrshort{ruwe}=1.22, which corresponds to $\alpha_{\rm \glsxtrshort{mse},\glsxtrshort{ruwe}}$=0.12\,mas with 2.8--$\sigma$ significance. Our analysis does not account for the presence of two planets. This situation will be explored in future studies. \\
    
    \item [\it USco 1621 A (\# 6250).] This young (5--10\,Myr) Upper Scorpius M2.5 star is known to host a very wide substellar companion with a mass of 15$\pm$2\,M$_{\rm J}$ at a projected separation of 2880$\pm$20\,au detected by direct imaging~\citep{Chinchilla2020}. The \texttt{ruwe}=1.176 of this source has a significance of 3.3--$\sigma$.  Our \glsxtrshort{pmex} map for USco 1621 A using the constraints from \glsxtrshort{ruwe} predicts a $>$5M$_{\rm J}$ companion within 10\,au of the star with 95.4\% confidence. The very wide companion is located beyond the 99.7\% confidence region, predicting a \glsxtrshort{ruwe} of smaller than 1.176. The \textit{Gaia} astrometry thus indicates the presence of a supplementary companion at smaller separation. As in the case of 2M0437 discussed above however, in such a young system one may expect significant accretion, which might induce astrometric jitter. \\
    
    \item [\it Transiting systems.] We find six systems with transiting planets only, namely K2-123~\citep{Livingston2018a} or \# 5975, K2-153~\citep{Livingston2018b} or \# 6669, K2-174~\citep{Barros2016,Livingston2019} or \# 1933, K2-321~\citep{Castro2020} or  \# 2832, Kepler-125 (2 planets;~\citealt{Rowe2014}) or \# 8550, and TOI-261~\citep{Hord2024} or \# 331. In all those systems, the planets are located within 0.1\,au and in the Neptunian regime in terms of radius, and thus also in terms of mass if applying a mass--radius relationship. However, our \glsxtrshort{pmex} maps, with the constraints from \glsxtrshort{ruwe} (1.25; 3.1--$\sigma$) and \glsxtrshort{pma} (0.087$\pm$0.053\,mas\,yr$^{-1}$; 0.9--$\sigma$) for TOI-261, predicts a mass of greater than 5M$_{\rm J}$ and at sma$<$20\,au ($<$3\,au for TOI-261) with 95.4\% confidence. In those systems, the \textit{Gaia} astrometry shows that additional companions exist at larger orbital periods. 
\end{itemize}

\begin{figure*}
    \centering
    \includegraphics[width=178.6mm,clip=true]{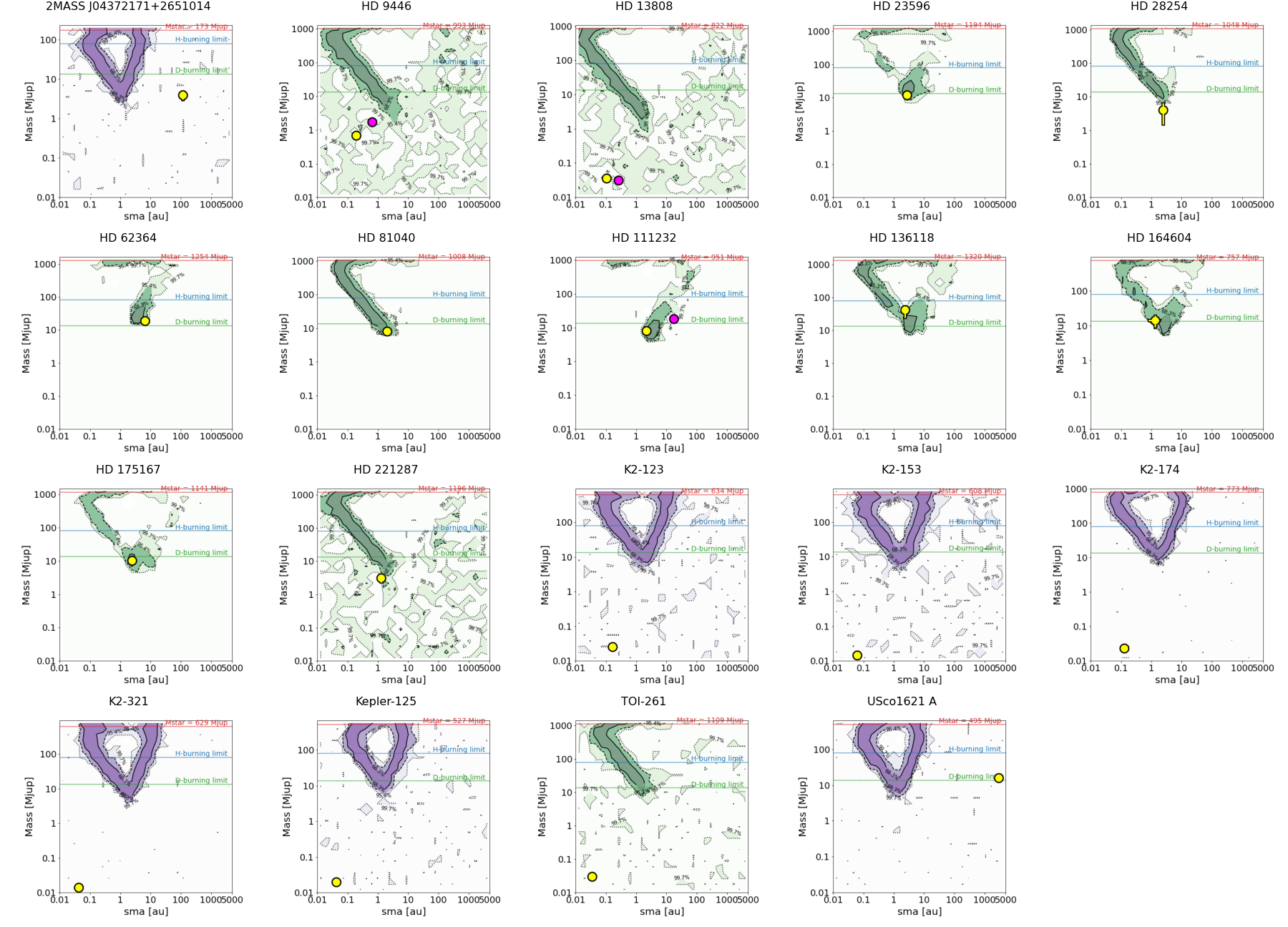}
    \caption{PMEX maps for the NEA cross-matched sample, combining, when possible, the constraints from \glsxtrshort{ruwe} and \glsxtrshort{pma} (green) or only \glsxtrshort{ruwe} (purple). The different shades of colour in each map are explained in Fig.~\ref{fig:GJ832_map}.}
    \label{fig:PMEX_BD_NEA}
\end{figure*}

\subsection{A focus on planet candidate hosts observed with \textsc{Hipparcos}}
\label{sec:xmatch_HIP}
We found 259 sources in our sample that also have a \textsc{Hipparcos} identifier. We used \glsxtrshort{pmex} on all the \glsxtrlong{hg} (\glsxtrshort{hg}) data of those 259 sources and found 132 sources with a proper motion astrometric signature, $\alpha_{\rm \glsxtrshort{pma}}$, more significant than 3--$\sigma$. We identified 20 sources with an \glsxtrshort{hg} astrometry that could be compatible with a planet companion, and 51 sources possibly compatible with a \glsxtrshort{bd} companion. For the other 62 sources, \glsxtrshort{pmex} predicts the existence of stellar companions. 

Among the 20 planet candidate hosts, we find 5 planet systems that were already discussed in Sect.~\ref{sec:xmatch_nea}: HD\,136118, HD\,111232, HD\,164604, HD\,175167, and HD\,23596. We find 9 spectroscopic binaries, including 7 SB1s, namely HD\,75767, HD\,17382, HD\,108510, HD\,112099, HD\,2085, HD\,23308, and HD\,221818, and 2 SB2s, HD\,30957 and BD+05\,3080. Among the remaining 8 systems, there is 1 system with a known planet or \glsxtrshort{bd} companion, HD\,40503, 1 candidate planetary system, HD\,33636, 2 wide visual binary systems, HD\,187129 and HD\,81697, and 2 systems with no known companions, CD-42 883 and HD\,105330. We discuss these eight cases below. 

\begin{itemize}[align = left]
    \item [\it HD\,40503 (\# 212).] This source is a K2/3V star located at 25.5\,pc. The astrometric orbit of HD\,40503 was characterized by \citet{Holl2023} and~\citet{Arenou2023}, and was compared with the fit of publicly available \glsxtrshort{rv} by~\citet{Marcussen2023}. Discrepant solutions were found: the \glsxtrshort{rv} implies a mass of 1.55\,M$_{\rm J}$ on an edge-on orbit with a period of 758 days, while astrometry implies a mass of 5.18$\pm$0.59\,M$_{\rm J}$ on an edge-on orbit with a period of 826$\pm$50\,days.  Figure~\ref{fig:PMEX_HD40503} shows the \glsxtrshort{pmex} confidence region of the mass and sma of the companion derived from the combination of \glsxtrshort{ruwe} (1.41; 4.2--$\sigma$) and \glsxtrshort{pma} (0.41$\pm$0.03\,mas\,yr$^{-1}$; $>$9--$\sigma$). Excluding the equal-mass binary scenario, the 68.3\% confidence region confirms that HD\,40503\,b could be a planet with an sma of 1.5--5\,au and a higher mass within 4--13.5\,M$_{\rm J}$. Our analysis of the five-parameter model residuals thus indicate that HD\,40503\,b must have a face-on orbit with an inclination of $<$3$^\circ$.
    \begin{figure}[hbt]
    \centering
    \includegraphics[width=89.3mm,clip=true]{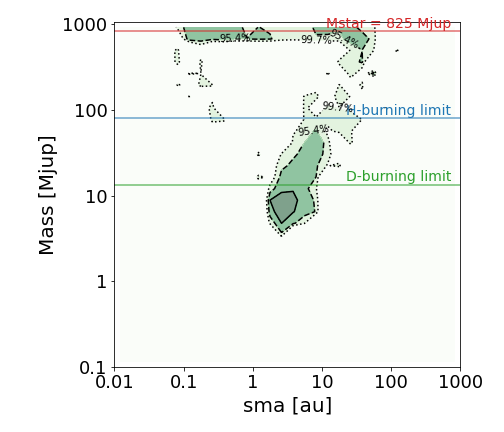}
    \caption{\glsxtrshort{pmex} maps for HD\,40503 based on the combination of constraints from \glsxtrshort{pma} and \glsxtrshort{ruwe}. Individual maps deduced from either \glsxtrshort{pma} or \glsxtrshort{ruwe} are shown in Fig.~E.1.}
    \label{fig:PMEX_HD40503}
    \end{figure} \\
    
    \item [\it HD\,33636 (\# 17).] This G0V star located at 29\,pc is known to host a planet candidate discovered by \glsxtrshort{rv} \citep{Perrier2003} and characterized with an $M\sin i$=9.28$\pm$0.77\,M$_{\rm J}$ and sma of 3.27$\pm$0.19\,au~\citep{Butler2006}. An analysis of the absolute astrometry of HD\,33636 using the Hubble Fine Guidance Sensor (HST-FGS), however, showed that the astrometric motion was rather compatible with a low-mass star with a mass of  142\,M$_{\rm J}$. Figure~\ref{fig:PMEX_HD33636} shows the \glsxtrshort{pmex} confidence region on the mass and sma of the companion derived from the combination of \glsxtrshort{ruwe} (1.88; 7.8--$\sigma$) and \glsxtrshort{pma} (0.34$\pm$0.04\,mas\,yr$^{-1}$; 4.9--$\sigma$). The 95.4\% confidence region predicts a mass of greater than 7\,M$_{\rm J}$ and an sma of smaller than 5\,au. Most importantly, at the location of the known companion, at about 3.3\,au, the \glsxtrshort{pmex} map excludes the possibility of a companion with a mass of greater than 40\,M$_{\rm J}$. This is, surprisingly, in total opposition to the result obtained from the HST-FGS astrometry. This is apparent in the individual maps obtained from considering either \glsxtrshort{ruwe} or \glsxtrshort{pma}. \citet{Xiao2023} already noted an inconsistency and proposed a smaller mass of 77.8$^{+6.9}_{-6.6}$\,M$_{\rm J}$  based on the combination of the \glsxtrshort{rv}
and the \glsxtrshort{hg} proper motion astrometry. Here, combining with the constraints from \glsxtrshort{ruwe}, we find a mass interval that is even lower and rather compatible with the initial value from~\citet{Butler2006}. Moreover, the value of the acceleration of HD\,33636 measured by \textit{Gaia} is published in the \glsxtrshort{nss} catalog, leading to $\gamma$=1.8\,mas\,yr$^{-2}$. Given that $\varpi$=34\,mas and an sma of 3.3\,au, we find that the star must be pulled by a companion of $\sim$15.4\,M$_{\rm J}$. HD\,33636\,b is thus a substellar companion at the planet--\glsxtrshort{bd} limit. 
    \begin{figure}[hbt]
    \centering
    \includegraphics[width=89.3mm,clip=true]{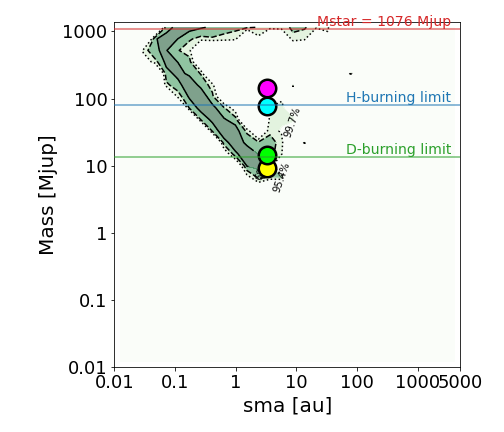}
    \caption{Same as Fig.~\ref{fig:PMEX_HD40503} but for HD\,33636. We added the four possible solutions: from \glsxtrshort{rv} only (yellow), \glsxtrshort{rv}+FGS (magenta), \glsxtrshort{rv}+\glsxtrshort{hg} (cyan), and \glsxtrshort{g3} acceleration (green).}
    \label{fig:PMEX_HD33636}
    \end{figure}   \\
    
    \item [\it HD\,187129 (\# 224).] This very wide visual binary system, also known as WDS\,19479+1002, and located at 100\,pc, is composed of two stars separated by $\sim$50\arcsec ($\sim$5050\,au) with a magnitude difference in the optical of $\Delta V$=1.47~\citep{Mason2024}. Our \glsxtrshort{pmex} map for HD\,187129, shown in Fig.~\ref{fig:PMEX_four_candidates}, combining the constraints from \glsxtrshort{ruwe} (1.20; 2.9--$\sigma$) and \glsxtrshort{pma} (0.21$\pm$0.03\,mas\,yr$^{-1}$; 4--$\sigma$) predicts a $>$4--M$_{\rm J}$ companion with an \glsxtrshort{sma}$<$100\,au with 95.4\% confidence. The main branch of the solution is located below 10\,au, with a smaller 68.3\% confidence region within 2--4\,au and 8--30\,M$_{\rm J}$, but other solutions at lower/larger sma and larger mass cannot be excluded with sufficient confidence.  \\
    
    \item [\it HD\,81697 (\# 205).] This wide visual binary system, also known as WDS\,09247-6055, and located at 67\,pc, is composed of two stars separated by $\sim$1.5\arcsec ($\sim$100\,au) with a magnitude difference in the optical of $\Delta V$=2.96~\citep{Mason2024}. Our \glsxtrshort{pmex} map for HD\,81697, shown in Fig.~\ref{fig:PMEX_four_candidates}, combining the constraints from \glsxtrshort{ruwe} (1.42; 4.2--$\sigma$) and \glsxtrshort{pma} (0.34$\pm$0.07\,mas\,yr$^{-1}$; 4.4--$\sigma$), predicts a $>$6--M$_{\rm J}$ companion with an \glsxtrshort{sma} of $<$10\,au with 95.4\% confidence. The 68.3\% confidence region is scattered but mainly centered within 1.5--6\,au and 8--30\,M$_{\rm J}$. Other solutions at lower and larger sma and larger mass, including the stellar companion at 100\,au, cannot be excluded with sufficient confidence.  \\

    \item [\it CD-42 883 (\# 422).] This star has a mass of 0.89\,M$_\odot$ and is thus possibly of G8 spectral type. To our knowledge, it is not a known binary or a planetary system. Combining the constraints from \glsxtrshort{ruwe} (1.23; 3.3--$\sigma$) and \glsxtrshort{pma} (0.48$\pm$0.04\,mas\,yr$^{-1}$; $>$9--$\sigma$), the \glsxtrshort{pmex} map for this source, shown in Fig.~\ref{fig:PMEX_four_candidates}, predicts at 95.4\% confidence that CD-42 883 has a companion of $>$10M$_{\rm J}$  at \glsxtrshort{sma}$>$2\,au. \\

    \item [\it HD\,105330 (\# 12).] This F8V star is known to show RV variability \citep{Nordstrom2004} but no orbit was ever determined for the possible companion in this system. Combining the constraints from \glsxtrshort{ruwe} (2.09; $>$9--$\sigma$) and \glsxtrshort{pma} (0.51$\pm$0.03\,mas\,yr$^{-1}$; 7.2--$\sigma$), the \glsxtrshort{pmex} map for this source, shown in Fig.~\ref{fig:PMEX_four_candidates}, predicts at 95.4\% confidence that HD\,105330 has a companion of $>$7--M$_{\rm J}$  at \glsxtrshort{sma}$<$10\,au. 
    
\end{itemize}

    \begin{figure*}[hbt]
    \centering
    \setlength{\unitlength}{1mm}
    \begin{picture}(178.6,160)
    \put(0,75){\includegraphics[width=89.3mm,clip=true]{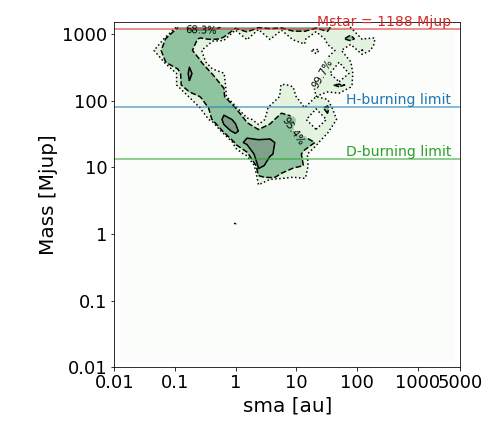}}
    \put(90,75){\includegraphics[width=89.3mm,clip=true]{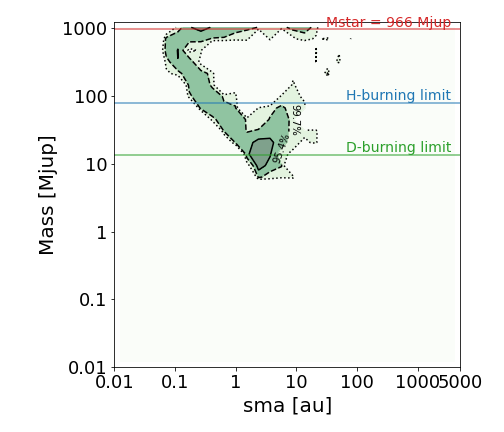}}
    \put(0,0){\includegraphics[width=89.3mm,clip=true]{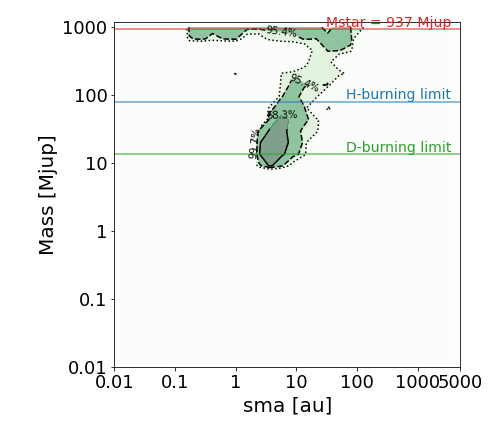}}
    \put(90,0){\includegraphics[width=89.3mm,clip=true]{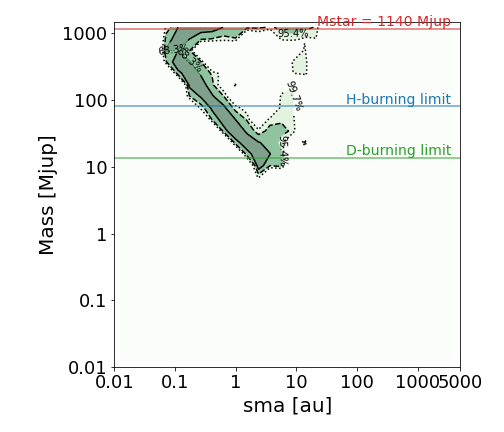}}
    \put(45,149){HD\,187129}
    \put(135,149){HD\,81697}
    \put(45,74){CD-42\,883}
    \put(131,74){HD\,105330}
    \end{picture}
    \caption{Same as Fig.~\ref{fig:PMEX_HD40503} but for HD\,187129 (top-left), HD\,81697 (top-right), CD-42\,883 (bottomn-left), and HD\,105330 (bottom-right). }
    \label{fig:PMEX_four_candidates}
    \end{figure*}
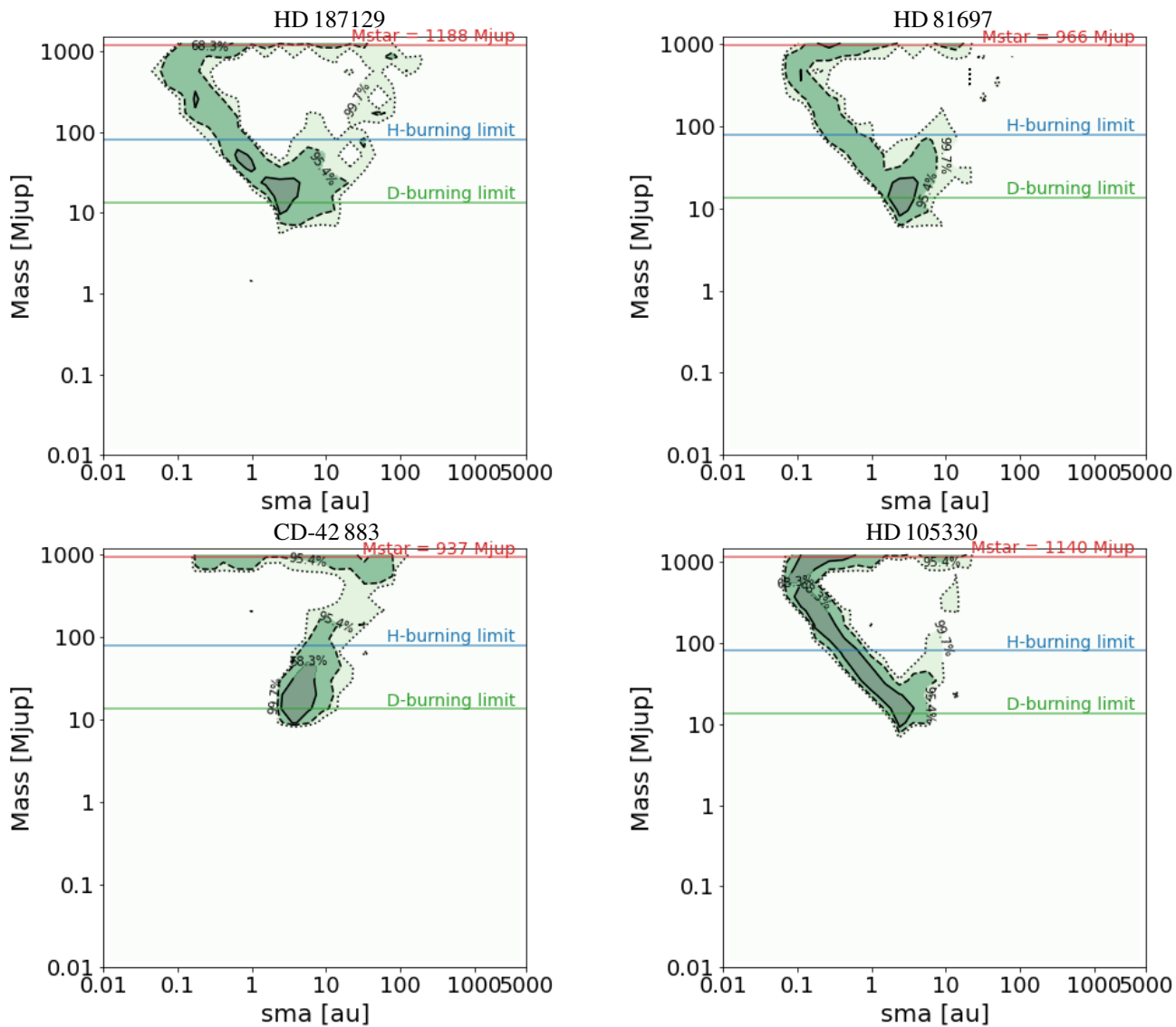

\noindent In summary, we  confirm five  planets (or low-mass \glsxtrshort{bd}) and find four new planet-candidate systems, HD\,187129, HD\,81697, CD-42 883, and HD\,105330. Among the subset of \textsc{Hipparcos} sources with a  $\alpha_{\rm \glsxtrshort{pma}}$ more significant than 3--$\sigma$ in our catalog, we find $\sim$39\% to be \glsxtrshort{bd} (or stellar) companions, $\sim$46\% to be binary stars, and 3.7--6.8\% to be planets. Extrapolating these percentages to our whole catalog suggests that most of the 9,698 systems that we identified do not contain planets, but rather \glsxtrshort{bd}s or stars. However, this extrapolation might not be permitted. For the \textsc{Hipparcos} sources for which both \glsxtrshort{ruwe} and \glsxtrshort{pma} are significant, the \glsxtrshort{pmex} map only leads to a 1--$\sigma$ confidence region upon the planetary companion at 1--3\,au when the individual maps from \glsxtrshort{ruwe} and \glsxtrshort{pma} almost fully overlap. Most values of \glsxtrshort{pma} would predict a larger mass for the companion at 1--3\,au and lead to 1--$\sigma$ confidence regions in the \glsxtrshort{bd} or stellar domain. Thus, the \textsc{Hipparcos} subsample might be strongly biased towards finding \glsxtrshort{bd}s or stellar companions. This implies that 3.7\% is a minimum planet rate among our catalog of 9,698 systems.

\section{Conclusion}

In Paper I, we introduced \glsxtrshort{pmex}, a tool that allows the characterization of the possible mass and sma of companions to stars observed with \textit{Gaia} using the \glsxtrlong{pma} (\glsxtrshort{pma}), the renormalized unit weight error (\glsxtrshort{ruwe}), and the \glsxtrlong{aen} (\glsxtrshort{aen}). \glsxtrshort{pmex} determines their significance within the null hypothesis that the star is single, and then modelizes them based on the star's reflex motion due to a companion, providing ranges of possible mass and \glsxtrshort{sma}. As mentioned in Sect.~\ref{sec:pmex}, the astrometric signature that is obtained from either the \glsxtrshort{ruwe} or the \glsxtrshort{aen} allows the determination of the minimum mass of a companion around any source of the \glsxtrshort{g3} database brighter than $G$=16.

In this work, we report an extensive catalog of 9,698 planet-candidate hosts with a primary star more massive than 0.5\,M$_\odot$ in which the astrometric signature of \glsxtrshort{ruwe}, $\alpha_{\rm \glsxtrshort{mse},\glsxtrshort{ruwe}}$, is more significant than 2.7--$\sigma$, predicting a minimum mass of a companion lying in the planetary domain, that is $<$13.5\,M$_{\rm J}$. Given the mass--\glsxtrshort{sma} degeneracy, many of these systems could actually be binaries, although a cross-match with existing catalogs of exoplanets and validated astrometric orbital solutions allowed us to confirm the planetary nature of some of our identified companions. A cross-check with the \glsxtrshort{nss} shows that our source selection process is approximately eight times more efficient at selecting planets than it is at selecting \glsxtrshort{bd} companions or binary stars in the domain of sensitivity of \textit{Gaia}.  Of the 260 systems observed with \textsc{Hipparcos,} focusing on the 134 systems that have a \glsxtrshort{pma} significance larger than 3--$\sigma$, we find that 5--9 of them (3.7--6.7\%) are likely detected planets. This could suggest that, nonetheless, at best $\sim$7\% of the sources of our catalog are truly planetary. However, this number was found using systems with a \glsxtrshort{pma} measurement, which favors binaries, unless both \glsxtrshort{pma} and \glsxtrshort{ruwe} coincide on predicting a planet companion at about 1--3\,au. A systematic vetting of this catalog will be carried out in future studies in order to determine the true frequency of binaries and planets in our sample of 9,698 sources. 

Finally, as shown in Paper I, \textit{Gaia}'s sensitivity is optimum for the detection of planets down to $\sim$0.1\,M$_{\rm J}$ around nearby ($<$10\,pc) low-mass ($<$0.5\,M$_{\odot}$) stars. We thus plan to extend this catalog to $M$-dwarfs with $M_\star$$<$0.5\,M$_\odot$ and $G>16$, which were excluded from the present version of the catalog.  

\section*{Data availability}
Appendices D--F are available at the following link at \href{https://zenodo.org/records/13734743?token=eyJhbGciOiJIUzUxMiJ9.eyJpZCI6ImMzZmU2MmEwLTJiNmUtNGYzMy05ZGIyLWVlYzEzOTM3YTZlNSIsImRhdGEiOnt9LCJyYW5kb20iOiJjMzNiNGU1ZTZkNDNhNmI3NjA1MWYwZjlhMzRjNzVlYSJ9.518w7Y5LIYsI9cGi1AxRyA9BOP9LVxFdzNnbqsTu1LEafM92uAfkBYwhoGGQGv3Q6WqaEe6mdYsa2KaquivenA}{https://zenodo.org}.
Table B.1 is only available in electronic form at the CDS via anonymous ftp to \url{cdsarc.u-strasbg.fr} (130.79.128.5) or via \url{http://cdsweb.u-strasbg.fr/cgi-bin/qcat?J/A+A/}.

\begin{acknowledgements}
We are very thankful to the anonymous referee for her/his thorough and courageous reading that led to significant improvements of this article. We thank D. S\'egransan for fruitful discussions. This work has made use of data from the European Space Agency (ESA) mission
{\it Gaia} (\url{https://www.cosmos.esa.int/gaia}), processed by the {\it Gaia}
Data Processing and Analysis Consortium (DPAC,
\url{https://www.cosmos.esa.int/web/gaia/dpac/consortium}). Funding for the DPAC
has been provided by national institutions, in particular the institutions
participating in the {\it Gaia} Multilateral Agreement. This work was granted access to the HPC resources of MesoPSL financed by the Region Ile de France and the project Equip\@Meso (reference ANR-10-EQPX-29-01) of the programme Investissements d’Avenir supervised by the Agence Nationale pour la Recherche. This project has received funding from the European Research Council (ERC) under the European Union's Horizon 2020 research and innovation programme (COBREX; grant agreement n° 885593). F.K. acknowledges funding from the initiative de recherches interdisciplinaires et stratégiques (IRIS) of Université PSL "Origines et Conditions d’Apparition de la Vie (OCAV)", as well as from the Action Pluriannuelle Incitative Exoplanètes from the Observatoire de Paris - Universit\'e PSL. F.K. also acknowledges funding from the American University of Paris.
\end{acknowledgements}

\bibliographystyle{aa}
\bibliography{main}

\clearpage
\onecolumn

\begin{appendix}

\section{Table of acronyms used in the text with their definitions and page references.}
\label{sec:acronyms}
\vspace{-1cm}
\setglossarystyle{longragged3col-booktabs}
\printglossary[type=\acronymtype,title=]
\newpage

\begin{landscape}
\section{Catalog of candidate systems with exoplanet}
\begin{table*}[hbt]
\caption{Extract of the full table of candidate systems with  the systems discussed in Sects.~\ref{sec:xmatch_nea} and~\ref{sec:xmatch_HIP}.}
\label{tab:planet_catalog}
\resizebox{\linewidth}{!}{%
\begin{tabular}{lccc|ccc|cc|ccc|ccccc|ccc|ccc|ccc|ccc}
\multicolumn{4}{l|}{Identifiers}             & \multicolumn{3}{c|}{Fluxes}                           & \multicolumn{2}{c|}{Colors}           & \multicolumn{3}{c|}{Stellar properties}  & \multicolumn{5}{c|}{\textit{Gaia} data}  & \multicolumn{3}{c|}{Noise estimations} & \multicolumn{3}{c|}{Astrometric signatures}  & \multicolumn{3}{c|}{Planet candidate properties} &   \multicolumn{3}{c}{Binary and planet crossmatch}                         \\
ID  &  \textit{Gaia} DR3 ID & main alias & HIP alias  & $G$ & \tablefootmark{$a$}$V$ & \tablefootmark{$a$}$K$ & \tablefootmark{$a$}$B-V$    & $Bp-Rp$ & \tablefootmark{$b$}$\varpi$ & \tablefootmark{$a$}sp type         & \tablefootmark{$c$}$M_{\star,{\rm Flame}}$ & \tablefootmark{$b$}dataset & \tablefootmark{$b$}$N_{\rm FoV}$ & \tablefootmark{$d$}$N_{\rm AL}$ & \tablefootmark{$b$}AEN & \tablefootmark{$b$}\texttt{ruwe}  & \tablefootmark{$e$}$\sigma_{\rm calib}$ & \tablefootmark{$e$}$\sigma_{\rm AL}$ & \tablefootmark{$e$}$\sigma_{\rm att}$  & \tablefootmark{$f$}$\alpha_{\rm UEVA,\texttt{ruwe}}$ & \tablefootmark{$f$}$p_{\alpha,{\rm \texttt{ruwe}}}$ & \tablefootmark{$f$}$s_{\alpha,{\rm \texttt{ruwe}}}$ & \tablefootmark{$g$}$M_{\rm p, min}$ & \tablefootmark{$g$}sma$_{\rm min}$ & \tablefootmark{$h$}$\rho_{\rm min}$  &  WDS alias & \tablefootmark{$i$}\glsxtrshort{nss}   & \tablefootmark{$j$}NEA      \\
    &              &            &            &     &                        &                        &                             &         &   (mas)                                   &                                    & (M$_\odot$)                                & 5p/6p                      &                                                  &                                 & (mas)                  &                                   & (mas)                                                              & (mas)                                & (mas)                                  & (mas)                                                   &                                                    &          ($N$--$\sigma$)                             & (M$_J$)                         & (au)                               & (mas)                                &            &         (0/1)            &     $N_{\rm planets}$       \\                                                                                                                
\hline
12 & 3472640938975300096 & HD 105330 & HIP 59135 & 6.59 & 6.73 & 5.30 & 0.510 & 0.706 & 30.49 & F8V & 1.088 & 5p & 51 & 9 & 0.363 & 2.087 & 0.164 & 0.049 & 0.077 & 0.305 & 0.00000 & $>$9 & 12.2 & 2.16 & 65.85 &  & 0 & 0 \\
16 & 4415515934099120768 & HD 136118 & HIP 74948 & 6.81 & 6.94 & 5.60 & 0.520 & 0.690 & 19.81 & F7V & 1.260 & 5p & 23 & 9 & 0.221 & 1.431 & 0.159 & 0.056 & 0.070 & 0.163 & 0.00164 & 3.1 & 11.1 & 2.27 & 44.94 &  & 0 & 1 \\
17 & 3238810137558836352 & HD  33636 & HIP 24205 & 6.86 &  & 5.57 &  & 0.750 & 33.80 & G0V\_CH-0.3 & 1.027 & 5p & 40 & 9 & 0.297 & 1.881 & 0.157 & 0.058 & 0.073 & 0.252 & 0.00000 & 7.5 & 8.7 & 2.12 & 71.61 &  & 1 & 0 \\
29 & 224870885460646016 & HD  23596 & HIP 17747 & 7.12 & 7.24 & 5.87 & 0.610 & 0.745 & 19.32 & F8 & 1.098 & 5p & 42 & 9 & 0.211 & 1.345 & 0.156 & 0.068 & 0.072 & 0.142 & 0.00071 & 3.4 & 9.0 & 2.17 & 41.87 & WDS J03480+4032A & 0 & 1 \\
30 & 5214209564293214208 & HD  62364 & HIP 36941 & 7.20 & 7.31 & 6.00 & 0.530 & 0.689 & 18.88 & F7V & 1.197 & 5p & 35 & 9 & 0.237 & 1.461 & 0.158 & 0.070 & 0.080 & 0.193 & 0.00000 & 4.8 & 13.2 & 2.23 & 42.10 &  & 0 & 2 \\
42 & 5855730584310531200 & HD 111232 & HIP 62534 & 7.42 & 7.61 & 5.90 & 0.680 & 0.879 & 34.61 & G8VFe-1.0 & 0.897 & 5p & 45 & 9 & 0.209 & 1.244 & 0.152 & 0.080 & 0.081 & 0.140 & 0.00036 & 3.6 & 4.3 & 2.03 & 70.09 &  & 1 & 2 \\
51 & 4781553628548774656 & HD  28254 & HIP 20606 & 7.51 & 7.68 & 6.01 & 0.755 & 0.889 & 18.12 & G5V & 1.000 & 5p & 32 & 9 & 0.235 & 1.509 & 0.152 & 0.083 & 0.077 & 0.193 & 0.00000 & 5.0 & 12.3 & 2.10 & 38.06 & WDS J04248-5037A & 0 & 1 \\
56 & 637329067477530368 & HD  81040 & HIP 46076 & 7.57 &  & 6.16 &  & 0.814 & 29.06 & G0V & 0.962 & 5p & 41 & 9 & 0.267 & 1.598 & 0.153 & 0.085 & 0.079 & 0.221 & 0.00000 & 6.7 & 8.5 & 2.07 & 60.25 &  & 1 & 1 \\
64 & 6489603813690967808 & HD 221287 & HIP 116084 & 7.69 & 7.81 & 6.57 & 0.500 & 0.663 & 17.87 & F7V & 1.142 & 5p & 58 & 9 & 0.157 & 1.223 & 0.160 & 0.088 & 0.074 & 0.123 & 0.00276 & 3.0 & 8.7 & 2.20 & 39.23 &  & 0 & 1 \\
74 & 6421118739093252224 & HD 175167 & HIP 93281 & 7.84 & 8.00 & 6.29 & 0.780 & 0.900 & 14.04 & G5IV/V & 1.089 & 5p & 73 & 9 & 0.181 & 1.173 & 0.150 & 0.092 & 0.074 & 0.108 & 0.00312 & 3.0 & 9.4 & 2.16 & 30.34 &  & 1 & 1 \\
89 & 4743692151804240896 & HD  13808 & HIP 10301 & 8.12 & 8.38 & 6.25 & 0.852 & 1.059 & 35.53 & K2V & 0.785 & 5p & 42 & 9 & 0.133 & 1.169 & 0.147 & 0.089 & 0.079 & 0.119 & 0.00487 & 2.8 & 3.3 & 1.94 & 68.82 &  & 0 & 2 \\
100 & 302956655074266112 & HD   9446 & HIP 7245 & 8.22 &  & 6.85 &  & 0.816 & 19.92 & G5V & 0.948 & 5p & 39 & 9 & 0.155 & 1.217 & 0.147 & 0.085 & 0.072 & 0.118 & 0.00639 & 2.7 & 6.6 & 2.06 & 41.09 &  & 0 & 2 \\
205 & 5299152644351701760 & HD  81697 & HIP 46151 & 8.92 & 9.46 & 7.28 & 0.330 & 0.925 & 15.04 & G8/K1 & 0.922 & 5p & 30 & 9 & 0.200 & 1.418 & 0.120 & 0.044 & 0.078 & 0.138 & 0.00004 & 4.1 & 10.0 & 2.04 & 30.74 & WDS J09247-6055AB & 0 & 0 \\
212 & 2884087104955208064 & HD  40503 & HIP 28193 & 8.97 & 9.21 & 7.03 & 0.990 & 1.124 & 25.51 & K2/3V & 0.788 & 5p & 39 & 9 & 0.157 & 1.408 & 0.116 & 0.040 & 0.076 & 0.122 & 0.00005 & 4.0 & 4.7 & 1.94 & 49.47 &  & 1 & 0 \\
224 & 4301963664605443328 & HD 187129 & HIP 97410 & 9.05 & 9.12 & 7.91 & 0.550 & 0.671 & 9.91 & G0 & 1.134 & 5p & 39 & 9 & 0.155 & 1.202 & 0.122 & 0.041 & 0.080 & 0.100 & 0.00494 & 2.8 & 12.7 & 2.19 & 21.71 & WDS J19479+1002B & 0 & 0 \\
302 & 4062446910648807168 & HD 164604 & HIP 88414 & 9.33 & 9.83 & 7.17 & 1.050 & 1.268 & 24.99 & K3.5Vk: & 0.745 & 5p & 28 & 9 & 0.177 & 1.158 & 0.110 & 0.032 & 0.128 & 0.174 & 0.00000 & 6.2 & 6.6 & 1.90 & 47.57 &  & 1 & 1 \\
331 & 2345033662373296768 & TOI-261 & HIP 4739 & 9.43 & 9.79 & 8.18 & 0.329 & 0.722 & 8.80 & F8/G0V & 1.059 & 5p & 57 & 9 & 0.127 & 1.253 & 0.119 & 0.036 & 0.075 & 0.091 & 0.00300 & 3.0 & 12.4 & 2.14 & 18.84 &  & 0 & 1 \\
422 & 4948630261544764032 & CD-42   883 & HIP 12368 & 9.76 & 9.94 & 8.19 & 0.740 & 0.902 & 12.92 &  & 0.894 & 5p & 57 & 9 & 0.136 & 1.234 & 0.116 & 0.035 & 0.077 & 0.094 & 0.00133 & 3.2 & 7.8 & 2.02 & 26.13 &  & 0 & 0 \\
1933 & 46217666333148416 & K2-174 &  & 12.00 &  & 9.50 &  & 1.488 & 9.98 & K7V & 0.738 & 5p & 56 & 9 & 0.106 & 1.430 & 0.105 & 0.019 & 0.073 & 0.091 & 0.00037 & 3.6 & 8.6 & 1.90 & 18.94 &  & 0 & 1 \\
2832 & 3856449627544923392 & K2-321 &  & 12.81 &  & 9.57 &  & 2.141 & 12.99 &  & 0.600 & 5p & 25 & 9 & 0.150 & 1.466 & 0.111 & 0.054 & 0.076 & 0.123 & 0.00024 & 3.7 & 7.8 & 1.77 & 23.01 &  & 0 & 1 \\
5975 & 685065155072248192 & K2-123 &  & 14.08 &  & 11.06 &  & 1.905 & 6.17 & M0V & 0.605 & 5p & 57 & 9 & 0.025 & 1.111 & 0.073 & 0.118 & 0.077 & 0.066 & 0.00321 & 2.9 & 8.9 & 1.78 & 10.96 &  & 0 & 1 \\
6250 & 6048608906890968960 & USco 1621 A &  & 14.15 &  & 10.19 &  & 2.676 & 7.38 & M2.5e & 0.660 & 5p & 50 & 9 & 0.073 & 1.176 & 0.079 & 0.120 & 0.075 & 0.078 & 0.00088 & 3.3 & 9.2 & 1.83 & 13.49 &  & 0 & 1 \\
6669 & 3700937760929835648 & K2-153 &  & 14.25 & 14.98 & 11.21 & 1.454 & 1.960 & 6.95 & M3.0V & 0.580 & 5p & 31 & 9 & 0.066 & 1.135 & 0.073 & 0.128 & 0.076 & 0.078 & 0.00570 & 2.8 & 9.0 & 1.75 & 12.17 &  & 0 & 1 \\
6883 & 151499478104075008 & 2MASS J04372171+2651014 &  & 14.31 &  & 10.39 &  & 2.860 & 7.81 & M4 & 0.642 & 5p & 27 & 8 & 0.089 & 1.214 & 0.084 & 0.129 & 0.074 & 0.096 & 0.00195 & 3.1 & 10.5 & 1.81 & 14.14 &  & 0 & 1 \\
8550 & 2086439488284337536 & Kepler-125 &  & 14.75 & 14.61 & 11.68 & 0.471 & 1.971 & 5.43 & M1V & 0.599 & 5p & 41 & 9 & 0.052 & 1.108 & 0.074 & 0.163 & 0.077 & 0.087 & 0.00374 & 2.9 & 13.1 & 1.77 & 9.61 &  & 0 & 2 \\
\hline
\end{tabular}%
}
\tablefoot{\\
\tablefoottext{$a$}{Taken from Simbad.}
\tablefoottext{$b$}{Taken from the \glsxtrshort{g3}.}
\tablefoottext{$c$}{CU8 database stellar mass, or \texttt{mass-Flame} (see Sect. \ref{sec:catalog}).}
\tablefoottext{$d$}{$N_{\rm AL}$ is the average number of along-scan angle measurements. See paper I for details.}
\tablefoottext{$e$}{Estimated from the \glsxtrshort{g3} catalog. See paper I for details.}
\tablefoottext{$f$}{$\alpha_{\rm UEVA, \texttt{ruwe}}$ is the astrometric signature calculated with \texttt{ruwe}, $p_{\alpha}$ is the confidence that $\alpha_{\rm UEVA}$ rejects the single-star hypothesis and $s_{\alpha}$ is the corresponding significance expressed as $N$-$\sigma$.}
\tablefoottext{$g$}{Mass and sma of the inferred companion at the minimum of the \glsxtrshort{pmex} \glsxtrshort{ruwe} curve. See paper I for details.}
\tablefoottext{$h$}{The separation $\rho={\rm sma}\times\varpi$.}
\tablefoottext{$i$}{Flag=1 if an orbital or acceleration solution is found in the \glsxtrshort{nss}, 0 otherwise.}
\tablefoottext{$j$}{Number of planets already detected in the system cross-matched from the NASA Exoplanet Catalog.} \\
The full catalog is available at the CDS
}
\end{table*}
\end{landscape}

\newpage
\twocolumn
\section{Details on the calculation of the predicted \texorpdfstring{$\alpha_{\rm UEVA}$}{astrometric signature} for the H23 sample}
\label{sec:details_holl2023}
We used the Eq.~\ref{eq:mass_sma_AEN} to calculate the $\alpha_{\rm \glsxtrshort{mse}}$ given the semi-major axis of a star and its orbital period $P$. There are two possible equations, one for $P$<3\,yr, with $\alpha_{\rm \glsxtrshort{mse}}$ directly proportional to the star's semi-major axis, and another for $P$>3\,yr, with $\alpha_{\rm \glsxtrshort{mse}}$ proportional to the gravitational pull due to the companion.
We used the semi-major axis of the photocenter divided by the parallax as an approximation of the semi-major axis of the primary star, $a_\star$, if not found in the \glsxtrshort{nss} catalog. The primary star's semi-major axis is known whenever there was, on top of astrometry, radial velocity data coming from the RVS instrument on-board, and an SB1/SB2 solution to the RV variations determined. 

Moreover, when the period is larger than 3\,yrs, if $M_\star$ was known and fulfilled the C$_3$ requirement of Sect.~\ref{sec:catalog}, we used the literature period given in H23's table A.1 to calculate the \glsxtrshort{sma} and mass of the companion by solving the equation based on Kepler's third law:
\begin{align}\label{eq:thirdlaw}
    {\rm sma}  = M_\star^{1/3}\,\left(1+q\right)^{1/3}\, \left(\frac{P}{\rm yr}\right)^{2/3}
\end{align}
with $q=M_c/M_{\star}$ the mass ratio. This can be expressed as a cubic equation on $Q=1/q$ since ${\rm sma}=a_\star\,(1+Q)$: 
\begin{align}
    Q^3 + 2\,Q^2 + Q - \frac{M_\star}{a_\star^3} \left(\frac{P}{\rm yr}\right)^{2} = 0 
\end{align}
which has a single real root given by: 
\begin{align}
    &\begin{cases}
a=1 \\
b=2 \\
c=1  \\
d= - M_\star\,a_\star^{-3} \,P({\rm yr})^{2} \\
\end{cases} \nonumber \\
&\begin{cases}
\Delta_0=b^2-3\,a\,c  \\
\Delta_1=2\,b^3-9\,a\,b\,c+27\,a^2\,d \\
\end{cases}   \nonumber  \\
&C={\rm sign}(\Delta_1)\sqrt[3]{\frac{|\Delta_1|+\sqrt{\Delta_1^2-3\,\Delta_0^3}}{2}}  \nonumber \\
&Q = -\frac{1}{3\,a}\,\left(b+C+\frac{\Delta_0}{C}\right)
\end{align}

From $Q$ and $M_\star$ we deduce $M_c$ and from Eq.~\ref{eq:thirdlaw} we find the relative \glsxtrshort{sma} of the companion to the primary star.  
The use of Eq.~\ref{eq:mass_sma_AEN} is then straightforward to determine a prediction of $\alpha_{\rm \glsxtrshort{mse}}$. Uncertainties are obtained by bootstrap, accounting for all input parameters and their individual uncertainties. All quantities entering this computation are written explicitly for all the 202 sources of H23 with $G$$<$16. 

\end{appendix}

\end{document}